\documentclass{emulateapj}
\usepackage{graphicx}
\usepackage{mathrsfs}	
\usepackage{amsmath}
\usepackage{color}
\usepackage{hyperref}
\usepackage{multirow}

\newcommand{\Survey}{BELLS GALLERY}

\shorttitle{BELLS IV.}
\shortauthors{Shu et al. 2016}
 
\begin{document}
 
\title{The BOSS Emission-Line Lens Survey. IV. : \\
Smooth Lens Models for the BELLS GALLERY Sample$^{\dag}$}

\altaffiltext{$^{\dag}$}{Based on observations made with the NASA/ESA Hubble Space Telescope, obtained from the Data Archive at the Space Telescope Science Institute, which is operated by AURA, Inc., under NASA contract NAS 5-26555. These observations are associated with program \#14189.}

\author{\mbox{Yiping Shu\altaffilmark{1}}}
\author{\mbox{Adam S. Bolton\altaffilmark{2, 3}}}
\author{\mbox{Shude Mao\altaffilmark{4, 1, 5}}}
\author{\mbox{Christopher S. Kochanek\altaffilmark{6}}}
\author{\mbox{Ismael P\'{e}rez-Fournon\altaffilmark{7, 8}}}
\author{\mbox{Masamune Oguri\altaffilmark{9, 10, 11}}}
\author{\mbox{Antonio D. Montero-Dorta\altaffilmark{2}}}
\author{\mbox{Matthew A. Cornachione\altaffilmark{2}}}
\author{\mbox{Rui Marques-Chaves\altaffilmark{7, 8}}}
\author{\mbox{Zheng Zheng\altaffilmark{2}}}
\author{\mbox{Joel R. Brownstein\altaffilmark{2}}}
\author{\mbox{Brice M\'{e}nard\altaffilmark{12}}}

\altaffiltext{1}{National Astronomical Observatories, Chinese Academy of Sciences, 20A Datun Road, Chaoyang District, Beijing 100012, China ({\tt yiping.shu@nao.cas.cn})}
\altaffiltext{2}{Department of Physics and Astronomy, University of Utah,
115 South 1400 East, Salt Lake City, UT 84112, USA}
\altaffiltext{3}{National Optical Astronomy Observatory, 950 N. Cherry Ave., Tucson, AZ 85719 USA}
\altaffiltext{4}{Physics Department and Tsinghua Centre for Astrophysics, Tsinghua University, Beijing 100084, China}
\altaffiltext{5}{Jodrell Bank Centre for Astrophysics, School of Physics and Astronomy, The University of Manchester, Oxford Road, Manchester M13 9PL, UK}
\altaffiltext{6}{Department of Astronomy \& Center for Cosmology and Astroparticle Physics, Ohio State University, Columbus, OH 43210, USA}
\altaffiltext{7}{Instituto de Astrof\'{i}sica de Canarias, C/ V\'{i}a L\'{a}ctea, s/n, 38205 San Crist\'{o}bal de La Laguna, Tenerife, Spain}
\altaffiltext{8}{Universidad de La Laguna, Dpto. Astrof\'{i}sica, E-38206 La Laguna, Tenerife, Spain}
\altaffiltext{9}{Research Center for the Early Universe, University of Tokyo, 7-3-1 Hongo, Bunkyo-ku, Tokyo 113-0033, Japan}
\altaffiltext{10}{Department of Physics, University of Tokyo, 7-3-1 Hongo, Bunkyo-ku, Tokyo 113-0033, Japan}
\altaffiltext{11}{Kavli Institute for the Physics and Mathematics of the Universe (Kavli IPMU, WPI), University of Tokyo, Chiba 277-8583, Japan}
\altaffiltext{12}{Department of Physics and Astronomy, Johns Hopkins University, Baltimore, MD 21218, USA}

\begin{abstract}

We present \textsl{Hubble Space Telescope} (\textsl{HST}) F606W-band imaging 
observations of 21 galaxy-Ly$\alpha$ emitter 
lens candidates in the Baryon Oscillation Spectroscopic Survey (BOSS) 
Emission-Line Lens Survey (BELLS) for the GALaxy-Ly$\alpha$ EmitteR sYstems (\Survey{}) 
survey. Seventeen systems are confirmed to be definite lenses with unambiguous evidence of 
multiple imaging. 
The lenses are primarily massive early-type galaxies (ETGs) at redshifts of 
approximately $0.55$, while the lensed sources are Ly$\alpha$ emitters (LAEs) 
at redshifts from two to three. Although most of the lens systems are well fit by 
smooth lens models consisting of singular isothermal ellipsoids in an 
external shear field, a thorough exploration of dark substructures in the lens 
galaxies is required. The Einstein radii of the \Survey{} lenses 
are, on average, $60\%$ larger than those of the BELLS lenses because of 
the much higher source redshifts. This will allow for 
a detailed investigation of the radius evolution of the mass profile in ETGs. 
With the aid of the average $\sim 13 \times$ lensing magnification, the LAEs 
are frequently resolved into individual star-forming knots with a wide range of 
properties. They have characteristic sizes from less than 100 pc to several 
kiloparsecs, rest-frame far-UV apparent AB magnitudes from 29.6 to 24.2, 
and typical projected separations of 500 pc to 2 kpc.

\end{abstract}


\slugcomment{Submitted to the ApJ}

\maketitle

\section{Introduction}

Strong gravitational lenses are sensitive to the total mass distribution regardless 
of its form. This allows for the extensive use of lenses as a powerful probe 
of lens galaxies including stars, dark matter, and dark substructures 
\citep[e.g.,][]{Kochanek95, Keeton98, Rusin03, SLACSV, SLACSVII, SLACSX, Vegetti10, 
Dutton11, Spiniello11, Vegetti12, Barnabe12, Bolton12, Brewer12, Brownstein12, 
Fadely12, MacLeod13, Sonnenfeld13, Oguri14, Nierenberg14, Shu15, Shu16a, 
Hezaveh16, Inoue16}. 
Furthermore, the magnification effect, by which the apparent size and total flux 
of the lensed source increase up to factors of tens, makes strong 
gravitational lensing a natural ``magnifier'' for studies of faint high-redshift 
objects \citep[e.g.,][]{Bolton06b, Quider09, Christensen12, Muzzin12, Bussmann13, 
Stark15, Karman16, Shu16a, Spilker16}.

\citet{Bolton04} developed a novel technique to efficiently select galaxy-scale 
strong gravitational lenses by searching along 
the line of sight (LOS) toward a potential lens for multiple emission lines 
from a common redshift beyond the redshift of the foreground object. 
Follow-up high-resolution \textsl{Hubble Space Telescope} 
(\textsl{HST}) imaging observations and associated lens models can then 
confirm the lensing nature of the systems. 
The application of this spectroscopic-selection and \textsl{HST}-observation 
strategy to the enormous database of galaxy spectra in the Sloan Digital Sky Survey 
\citep[SDSS;][]{York00} and the Baryon Oscillation Spectroscopic Survey 
\citep[BOSS;][]{Dawson13} of the Sloan Digital Sky Survey-III 
\citep[SDSS-III;][]{Eisenstein11} has resulted in four dedicated surveys: 
the Sloan Lens ACS \citep[SLACS;][]{SLACSI} survey, 
the Sloan WFC Edge-on Late-type Lens Survey \citep[SWELLS;][]{SWELLSI}, 
the SLACS for the Masses \citep[S4TM;][]{Shu15} survey, 
and the BOSS Emission-Line Lens Survey \citep[BELLS;][]{Brownstein12}. 
Over $150$ grade-A galaxy lenses with clear and convincing evidence for multiple 
imaging have been discovered from the four surveys leading to a broad range 
of scientific discoveries \citep[e.g.,][]{SLACSII, SLACSIII, SLACSIV, Czoske08, 
SLACSV, SLACSVI, SLACSVII, SLACSVIII, Barnabe09, SLACSIX, SLACSX, Grillo10, 
SLACSXI, Barnabe11, Dutton11, Barnabe12, Bolton12, Brewer12, Dutton13, Brewer14, 
Shu15}. 

We initiated the BELLS for GAlaxy-Ly$\alpha$ EmitteR sYstems (BELLS GALLERY) survey 
in 2015 to search for low-mass, dark substructures in lens galaxies by utilizing the 
intrinsic compactness of high-redshift lensed Ly$\alpha$ emitters (LAEs). 
The galaxy-LAE strong lens candidate systems are spectroscopically selected from 
almost 1.5 million galaxy spectra in the final data release (DR12) of 
the BOSS survey. Cuts on the detection significance and apparent flux and 
profile of the detected emission line were applied to the parent sample of 
187 candidates to select the 21 highest-quality candidates 
that compose the \Survey{} sample.  
The foreground lenses are classified as massive early-type galaxies (ETGs) with a 
median redshift of 0.55 as determined by the BOSS spectroscopic data. 
The background sources are LAEs at redshifts from 2 to 3 with 
a median redshift of 2.5. 
Detailed descriptions of the selection algorithm and properties of the sample 
are presented in a previous paper \citep[BELLS-III;][]{BELLSIII}. 

This paper, as the fourth in a series about the BELLS survey, presents the follow-up 
imaging data from the recently finished \textsl{HST} observations along with 
lens models constructed from the data. It is organized as follows. 
Section 2 provides the \textsl{HST} observational data and the derived 
photometric properties of the lens galaxies. 
Smooth lens models (without any substructure) are presented in Section 3. 
Discussion and conclusions are given in Sections 4 and 5, respectively. 
Throughout the paper, 
we adopt a fiducial cosmological model with $\rm \Omega_m = 0.274$, 
$\rm \Omega_{\Lambda} = 0.726$, and $H_0 \rm = 70\,km\,s^{-1}\,Mpc^{-1}$ 
\citep[WMAP7,][]{WMAP7}.

\begin{figure*}[htbp]
\centering
\includegraphics[width=0.99\textwidth]{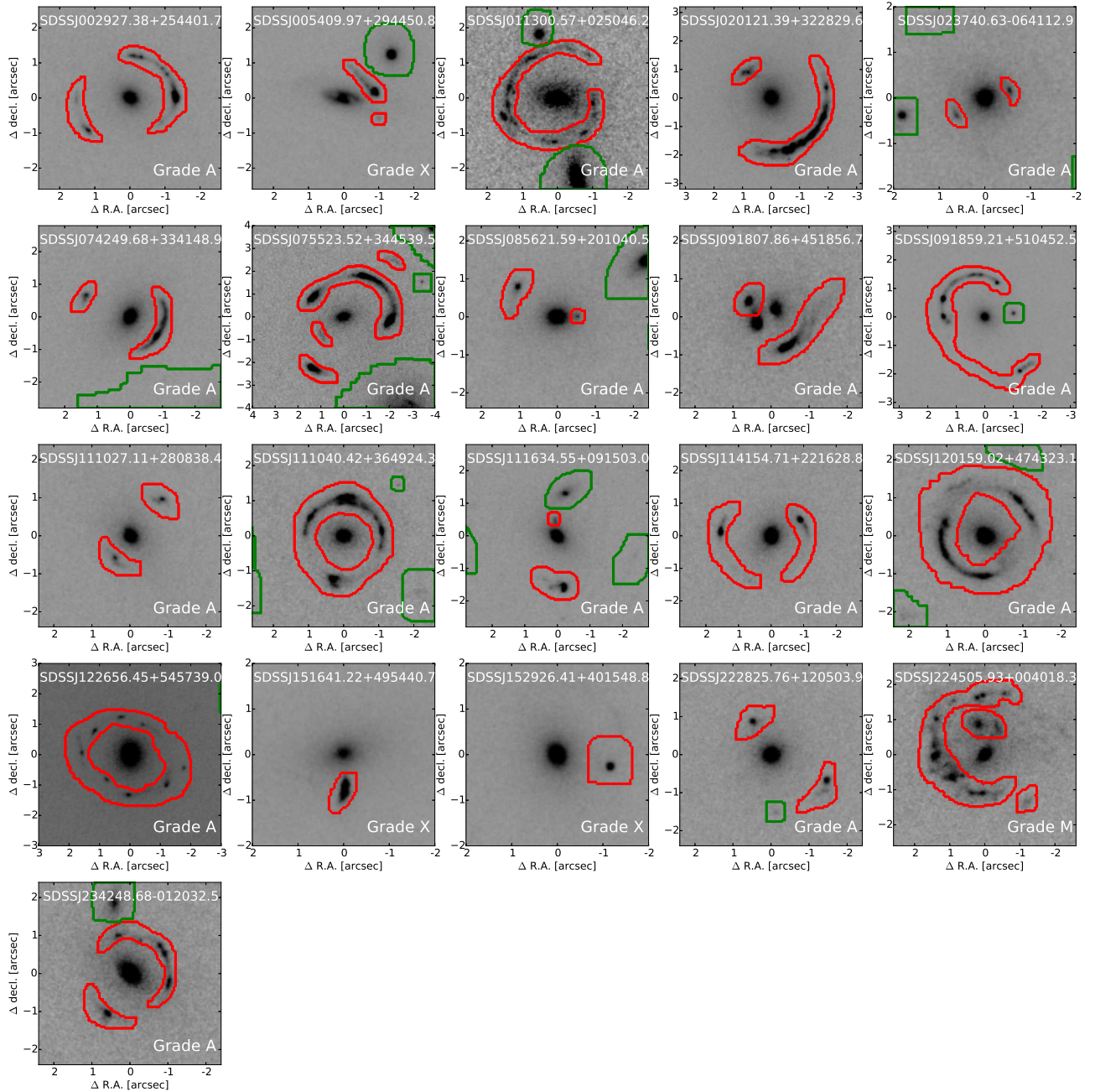}
\caption{\label{fig:mosaic}
Mosaic of the \textsl{HST} F606W-band images of 21 \Survey{} candidate systems. 
The images are orientated such that 
north is up and east is to the left. The axis labels provide offsets in R.A. 
and decl., respectively. The regions bounded by the red dots are regions 
that are thought to be related to the background source, while the green dots 
outline regions with contaminating structures. 
All systems but J005409.97$+$294450.8, J151641.22$+$495440.7, 
J152926.41$+$401548.8, and J224505.93$+$004018.3 are clearly multiply imaged 
lenses (Grade A). 
J005409.97$+$294450.8, J151641.22$+$495440.7, and J152926.41$+$401548.8 are 
only singly imaged (Grade X), while J224505.93$+$004018.3 is too complex to classify 
(Grade M). 
}
\end{figure*}

\section{\textsl{HST} Data}

The \textsl{HST} follow-up observations of the 21 \Survey candidates 
started in 2015 November under \textsl{HST} Cycle 23 program ID 
14189 (PI: A. Bolton) and finished in 2016 May. 
For each candidate, 4 sub-exposures of approximately 630 s each 
were taken within a single \textsl{HST} orbit 
using the \emph{V}-band F606W filter on the Wide Field Camera 3 (WFC3). 
The individual flat-fielded (FLT) sub-exposure files for each system 
were fully reduced, rectified onto uniform pixels of 0\farcs04, 
and combined using the custom-built and extensively tested 
reduction pipeline, {\tt ACSPROC}, following the procedures described in 
\citet{SLACSV}, \citet{Brownstein12}, and \citet{Shu15}. 
We modified {\tt ACSPROC} for the shift from the ACS/WFC to the 
WFC3/UVIS camera. 
The empirical point spread function (PSF) was generated by the {\tt Tiny Tim} tool 
\citep{Krist93}. Pixel count errors are rescaled following 
\citet{Shu15, Shu16a} to correct for possible error correlations created by 
imaging resampling during the data reduction process. 
The mean pixel count error after the rescaling is $\sim 0.005$ electrons per second per pixel$^2$.

\begin{table*}[htbp]
\begin{center}
\caption{\label{tb:tb1} Selected properties of the BELLS GALLERY sample.}
\begin{tabular}{c c c c c c c c c c}
\hline \hline
Target & \multicolumn{4}{c}{BOSS} & & \multicolumn{3}{c}{\textsl{HST} F606W} & Classification \\ 
\cline{2-5} \cline{7-9}
 & $z_{L}$ & $z_{S}$ & $\sigma_{\rm BOSS}$ & $R_{\rm dev}$ & & $R_{\rm eff}$ & $q$ & P.A. & \\
 & & & (km\,s$^{-1}$) & (arcsec) & & (arcsec) & & (degree) & \\
(1) & (2) & (3) & (4) & (5) & & (6) & (7) & (8) & (9) \\
\hline
SDSSJ002927.38$+$254401.7 & 0.5869 & 2.4504 &  241 $\pm$   45 & 1.43 $\pm$       0.71 & & 0.49 $\pm$ 0.01 & 0.825 $\pm$      0.006 & \phantom{1}52.2 $\pm$ 1.1 & E-S-A \\ 
\hline
SDSSJ005409.97$+$294450.8 & 0.4488 & 2.7176 &  177 $\pm$   58 & 1.31 $\pm$       0.37 & & 0.39 $\pm$ 0.01 & 0.457 $\pm$      0.007 & \phantom{1}81.6 $\pm$ 0.5 & L-S-X \\ 
\hline
SDSSJ011300.57$+$025046.2 & 0.6230 & 2.6088 &  850 $\pm$   -1 & 2.80 $\pm$       1.26 & & 1.84 $\pm$ 0.22 & 0.645 $\pm$      0.008 & \phantom{1}79.3 $\pm$ 0.8 & E-M-A \\ 
\hline
SDSSJ020121.39$+$322829.6 & 0.3957 & 2.8209 &  256 $\pm$   20 & 2.60 $\pm$       0.38 & & 2.32 $\pm$ 0.16 & 0.882 $\pm$      0.004 & \phantom{1}21.1 $\pm$ 1.1 & E-S-A \\ 
\hline
SDSSJ023740.63$-$064112.9 & 0.4859 & 2.2491 &  290 $\pm$   89 & 1.05 $\pm$       0.45 & & 1.05 $\pm$ 0.09 & 0.980 $\pm$      0.008 & \phantom{1}109.0 $\pm$ 11.1 & E-S-A \\ 
\hline
SDSSJ074249.68$+$334148.9 & 0.4936 & 2.3633 &  218 $\pm$   28 & 1.07 $\pm$       0.32 & & 0.89 $\pm$ 0.03 & 0.717 $\pm$      0.004 &  148.3 $\pm$ 0.4 & E-S-A \\ 
\hline
SDSSJ075523.52$+$344539.5 & 0.7224 & 2.6347 &  272 $\pm$   52 & 0.27 $\pm$       0.72 & & 2.89 $\pm$ 0.47 & 0.602 $\pm$      0.007 &  102.4 $\pm$ 0.6 & E-S-A \\ 
\hline
SDSSJ085621.59$+$201040.5 & 0.5074 & 2.2335 &  334 $\pm$   54 & 1.15 $\pm$       0.42 & & 0.51 $\pm$ 0.01 & 0.795 $\pm$      0.005 & \phantom{1}91.9 $\pm$ 0.8 & E-S-A \\ 
\hline
SDSSJ091807.86$+$451856.7 & 0.5238 & 2.3440 &  119 $\pm$   61 & 2.08 $\pm$       1.16 & & 2.22 $\pm$ 0.73 & 0.807 $\pm$      0.016 & \phantom{1}58.6 $\pm$ 2.5 & E-M-A \\ 
\hline
SDSSJ091859.21$+$510452.5 & 0.5811 & 2.4030 &  298 $\pm$   49 & 1.89 $\pm$       0.77 & & 0.57 $\pm$ 0.02 & 0.880 $\pm$      0.008 & \phantom{1}39.1 $\pm$ 2.1 & E-S-A \\ 
\hline
SDSSJ111027.11$+$280838.4 & 0.6073 & 2.3999 &  191 $\pm$   39 & 0.40 $\pm$       0.23 & & 1.45 $\pm$ 0.16 & 0.762 $\pm$      0.006 & \phantom{1}31.4 $\pm$ 0.8 & E-S-A \\ 
\hline
SDSSJ111040.42$+$364924.4 & 0.7330 & 2.5024 & \,\,\,531 $\pm$  165 & 0.88 $\pm$       0.40 & & 0.39 $\pm$ 0.01 & 0.779 $\pm$      0.008 & \phantom{1}88.4 $\pm$ 1.1 & E-S-A \\ 
\hline
SDSSJ111634.55$+$091503.0 & 0.5501 & 2.4536 &  274 $\pm$   55 & 0.95 $\pm$       0.35 & & 0.98 $\pm$ 0.06 & 0.690 $\pm$      0.005 & \phantom{1}41.0 $\pm$ 0.5 & E-S-A \\ 
\hline
SDSSJ114154.71$+$221628.8 & 0.5858 & 2.7624 &  285 $\pm$   44 & 0.63 $\pm$       0.25 & & 0.44 $\pm$ 0.01 & 0.801 $\pm$      0.005 &  157.6 $\pm$ 0.9 & E-S-A \\ 
\hline
SDSSJ120159.02$+$474323.2 & 0.5628 & 2.1258 &  239 $\pm$   43 & 1.83 $\pm$       0.45 & & 0.48 $\pm$ 0.01 & 0.738 $\pm$      0.006 & \phantom{1}57.9 $\pm$ 0.7 & E-S-A \\ 
\hline
SDSSJ122656.45$+$545739.0 & 0.4980 & 2.7322 &  248 $\pm$   26 & 1.12 $\pm$       0.20 & & 0.56 $\pm$ 0.01 & 0.827 $\pm$      0.003 & \phantom{11}2.0 $\pm$ 0.6 & E-S-A \\ 
\hline
SDSSJ151641.22$+$495440.7 & 0.5479 & 2.8723 &  226 $\pm$   40 & 1.54 $\pm$       0.39 & & 1.71 $\pm$ 0.07 & 0.618 $\pm$      0.004 &  107.7 $\pm$ 0.3 & E-S-X \\ 
\hline
SDSSJ152926.41$+$401548.8 & 0.5308 & 2.7920 &  283 $\pm$   33 & 1.56 $\pm$       0.52 & & 1.65 $\pm$ 0.11 & 0.798 $\pm$      0.004 & \phantom{1}18.6 $\pm$ 0.7 & E-S-X \\ 
\hline
SDSSJ222825.76$+$120503.9 & 0.5305 & 2.8324 &  255 $\pm$   50 & 0.82 $\pm$       0.39 & & 0.53 $\pm$ 0.02 & 0.951 $\pm$      0.007 &  106.3 $\pm$ 4.3 & E-S-A \\ 
\hline
SDSSJ224505.93$+$004018.3 & 0.7021 & 2.5413 & \,\,\,64 $\pm$   44 & 1.84 $\pm$       0.65 & & 2.69 $\pm$ 0.51 & 0.687 $\pm$      0.009 &  125.5 $\pm$ 1.0 & E-S-M \\ 
\hline
SDSSJ234248.68$-$012032.5 & 0.5270 & 2.2649 &  274 $\pm$   43 & 1.32 $\pm$       0.66 & & 1.75 $\pm$ 0.19 & 0.682 $\pm$      0.006 & \phantom{1}44.9 $\pm$ 0.6 & E-S-A \\ 
\hline \hline
\end{tabular}
\end{center}
\textsc{      Note.} --- Column 1 is the SDSS system name in terms of the truncated J2000 R.A. and decl. in the format HHMMSS.ss$\pm$DDMMSS.s. Columns 2 and 3 are the redshifts of the foreground lens and the background LAE inferred from the BOSS spectrum. Column 4 is the velocity dispersion calculated from the velocity-dispersion likelihood function using restricted stellar eigenspectra with redshift-error marginalization, as described in \citet{Shu12}. The best-fit velocity dispersion for SDSS\,J011300.57$+$025046.2, 850 km\,s$^{-1}$, is at the maximum dispersion value tested and is therefore unreliable. The associated error is set to -1 as a warning flag. Column 5 is the BOSS $r$-band de Vaucouleurs fit effective radius (in the intermediate axis convention) of the foreground lens. Columns 6-8 are the effective radius (in the intermediate axis convention), minor-to-major axis ratio, and major-axis position angle of the lens galaxy with respect to the north inferred from \textsl{HST} F606W-band imaging data assuming a S\'{e}rsic model. For systems with multiple lenses, values for the primary lenses are reported. Column 9 is the classification with codes denoting the foreground-lens morphology, the foreground-lens multiplicity, and the status of system as a lens based on available data. Morphology is coded by ``E'' for early-type (elliptical and S0) and ``L'' for late-type (Sa and later). Multiplicity is coded by ``S'' for single and ``M'' for multiple. Lens status is coded by ``A'' for systems with clear and convincing evidence of multiple imaging, ``M'' for systems with possible evidence of multiple imaging, and ``X'' for non-lenses. \\
\end{table*}

Figure~\ref{fig:mosaic} shows a mosaic of the fully reduced images of the 21 
candidate systems in the \Survey{} survey. The cutouts are centered on the 
R.A. and decl. of the lens galaxy as determined by the BOSS survey. 
For all systems but J091859.21$+$510452.5 
(the ``SDSS'' is omitted to save the space), 
the cutout centers are coincident with the 
lens galaxy centers. The two lens components in J091859.21$+$510452.5 are 
not resolved by BOSS, and therefore its cutout is centered between the two lens 
components. 
For the background LAEs, the \textsl{HST} F606W filter covers 
their rest-frame far ultraviolet (UV) emission.
We manually generate feature 
masks that enclose the regions that are suspected to be lensed features (red dots). 
Junk masks for contaminating structures are also generated 
in the same fashion and outlined by the green dots. 
The pixel count errors within the junk masks are set to infinity 
so that they do not affect the actual fitting. 
To be compatible with the previous SLACS, BELLS, and S4TM surveys, we carry out 
a smooth b-spline fit to the foreground-light distribution of each system 
following \citet{SLACSI, SLACSV}. 
Besides the b-spline model, an elliptical S\'{e}rsic model 
\citep{Sersic63} is also used for the foreground-light subtraction. 
As will be explained in the following section, the S\'{e}rsic fit is performed 
simultaneously with the lens modeling to reduce potential systematics.  
The b-spline-subtracted residual image of every candidate system is inspected 
visually for lensed features, and a classification code, which 
characterizes the lens morphology, multiplicity, and status, 
is assigned accordingly. 

Table~\ref{tb:tb1} summarizes the properties of the \Survey{} sample measured 
by the BOSS data reduction pipeline \citep{Bolton12b}, 
the S\'{e}rsic fit parameters based on the \textsl{HST} imaging data, 
and the classification results. 
Although not directly comparable because of differences in 
data quality and assumed light profile, the effective radii measured from the BOSS 
and HST data are in general agreement. The axis ratio distribution measured from 
HST data is consistent with that for ETGs in general \citep[e.g.,][]{Hao06}. 
In terms of lens status, 
17 of the 21 candidate systems are confirmed to be grade-A galaxy-LAE lenses 
with unambiguous lensed features. This includes 8 lenses with extended arcs, 
3 with quadruple images, and 6 with double images. 
Three systems, J005409.97$+$294450.8, J151641.22$+$495440.7, 
and J152926.41$+$401548.8, are simply non-lenses with singly imaged sources. 
The remaining system, J224505.93$+$004018.3, shows complex structures that 
are too hard to interpret based on this single-band imaging data, 
so we conservatively consider this system a ``maybe'' for the moment. 
Further multi-band imaging data might reveal the true status of this system. 
The success rate of 81\% (17 lenses from 21 candidates) is much higher than those 
for previous SLACS, BELLS, and S4TM surveys based on similar selection techniques. 
This is presumably due to the much more 
stringent selection cuts applied to the \Survey{} sample and the 
inclusion of two systems, J020121.39$+$322829.6 and J075523.52$+$344539.5, that 
do not meet the selection thresholds but show definite evidence for strong-lensing 
features in their color-composite SDSS images. 
There remain another 166 candidate systems in the parent sample with 
high signal-to-noise ratio (SNR) detections of ``anomalous'' emission lines. 
Considering the typical 50\% success rate in the previous surveys, we expect 
another $\sim 70$ galaxy-LAE lens systems from future follow-up observations of 
the remaining sample. 

\section{Smooth Lens Models}

\subsection{Methodology}

Building on our previous works \citep{SLACSV, Brownstein12, Shu15, Shu16a}, 
we develop an open source lens modeling tool {\tt lfit\_gui} with a graphical 
user interface (GUI). In this subsection, we use the \Survey{} lenses to 
demonstrate the settings and functions of {\tt lfit\_gui}. 

\subsubsection{Foreground-light Subtraction}

Foreground-light removal is a crucial step in gravitational lens modeling 
when the emission from the foreground lens is high enough to hinder the correct 
interpretation of the lensed emission. Being massive elliptical galaxies, 
the \Survey{} lenses contribute substantial fractions of the observed 
optical emission. One commonly used approach is 
to perform the foreground-light subtraction separately prior to the lens modeling 
\citep[e.g.,][]{Moustakas07, SLACSV, Suyu10, SLACSXI, Brownstein12, 
Sonnenfeld13, Shu15}. However, this can lead to two problems: 
(1) automatic adjustments are not possible if the lensing and junk features are 
masked inappropriately, which can affect the foreground-light fitting; 
(2) the pixel count errors estimated for the full data might not be appropriate 
for the foreground-subtracted residual, and hence bias the fitting process. 
Therefore, we perform the foreground-light subtraction jointly 
with the lens modeling, as done in \citet{Shu16a} and \citet{Rusu16}. 

In particular, we use the elliptical S\'{e}rsic profile to model 
the foreground light in the joint modeling. 
Note that {\tt lfit\_gui} also offers two other commonly used profiles, 
the core-S\'{e}rsic profile \citep{Graham03, Trujillo04} and the Hernquist profile 
\citep{Hernquist90}, as models for the foreground light. 
To investigate the impact of the joint modeling approach, 
we perform lens modeling on the b-spline-subtracted residual image as well. 
The results are compared in Section~\ref{sect:results}. 

\subsubsection{Lens Mass Model}

Following our previous works \citep{SLACSV, Brownstein12, Shu15, Shu16a}, 
the mass distribution of 
the foreground lens is modeled with the singular isothermal ellipsoid (SIE) model, 
which has a projected two-dimensional surface mass density profile of 
\begin{equation}
\Sigma (x, y) = \Sigma_{\rm crit} \frac{\sqrt{q}}{2} \frac{b_{\rm SIE}}{\sqrt{x^2+q^2 y^2}},
\end{equation}
where $\Sigma_{\rm crit}$ is the critical density determined by the cosmological 
distances as 
\begin{equation}
\Sigma_{\rm crit} = \frac{c^2}{4 \pi G} \frac{d_{S}}{d_{L} d_{LS}}, 
\end{equation}
and $d_{L}$, $d_{S}$, and $d_{LS}$ are the angular diameter distances 
from the observer to the lens, from the observer to the source, 
and between the lens and the source, respectively. 
The lensing strength of the SIE model $b_{\rm SIE}$ is equivalent to the 
Einstein radius in the ``intermediate axis'' convention for elliptical models, 
and $q$ is the minor-to-major axis ratio of the isodensity contours.  
We use one or two SIE components depending on the foreground-lens multiplicity. 

We further include an external shear for three systems residing in crowded 
environments for which the pure SIE models fail to yield good fits. 
The effective lensing potential of the external shear is  
\begin{equation}
\psi_{\rm shear} (r, \phi) = -\frac{\gamma}{2} r^2 \cos 2(\phi-\phi_{\gamma}), 
\end{equation}
in which $\gamma$ and $\phi_{\gamma}$ are the strength and position angle of 
the external shear. Note that singular power-law ellipsoid \citep{Tessore15}, 
spherical Navarro-Frenk-White \citep{Bartelmann96, Golse02}, 
and point mass lens models are also provided in {\tt lfit\_gui}. 

\subsubsection{Source-light Model}

The surface brightness distribution of the background source is reconstructed  
parametrically using elliptical S\'{e}rsic models. 
Because the morphologies of the lensed LAEs are  
highly irregular and clumpy, multiple S\'{e}rsic components are 
usually needed to recover the observed lensing features. 
We therefore employ a pixelized source model as a guide to determine 
the number of required S\'{e}rsic components. We obtain a first guess of 
the model parameters by considering only one S\'{e}rsic component, which is 
able to capture the most significant features in the observational data. 
We then keep the lens model parameters fixed and generate a pixelized source 
model based on which extra S\'{e}rsic components are added. 
This is done iteratively until the parametric source model and the pixelized 
source model are in reasonable agreement. 

The pixelized source model is obtained through the semilinear inversion method 
first introduced by \citet{Warren03}. 
Following the later development of this technique 
\citep{Dye05, Koopmans05, Brewer06, Suyu06, Vegetti09, Nightingale15}, 
we adopt an irregular source grid constructed from the Delaunay tessellation for 
a set of $N$ points in the source plane that are directly mapped from $N$ pixels 
in the image plane through the lens equation given a particular lens model. 
The convolution with the PSF is incorporated as part of the process. 
The $N$ pixels in the image plane are chosen to contain the $N_b$ pixels 
on the boundaries of the feature masks and the brightest $N-N_b$ pixels within 
the feature masks. The number $N$ is chosen to be equal to half of the total pixel 
number within the feature masks. The inclusion of the $N_b$ boundary pixels ensures 
that any pixel within the feature masks will be mapped inside the 
Delaunay tessellation. We choose to use the brightest $N-N_b$ pixels in the 
image plane to preserve as many of the high significance features as possible. 
Additional linear regularization is included to regulate the overall smoothness of 
the pixelized source \citep{Warren03} . 

\begin{figure}[htbp]
\includegraphics[width=0.49\textwidth]{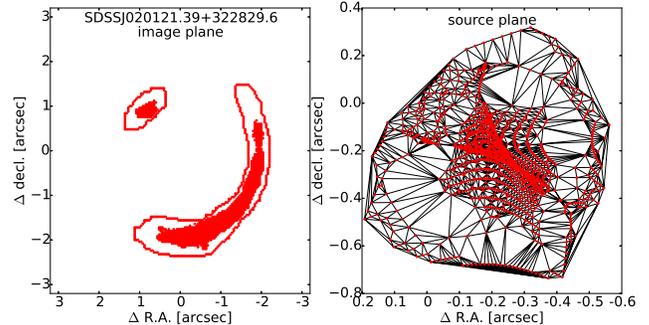}
\caption{\label{fig:illustration}
Illustration of the irregular source grid construction from a Delaunay tessellation 
for the grade-A lens J020121.39$+$322829.6. 
The left panel shows the pixels (red dots) 
in the image plane that are used to construct the Delaunay tessellation, while 
the right panel shows the corresponding positions (red dots) in the source plane 
after the mapping. The black lines outline the resulting Delaunay triangles.}
\end{figure}

Figure~\ref{fig:illustration} illustrates this source gridding process using 
the grade-A lens system J020121.39$+$322829.6 as an example. 
The feature masks for J020121.39$+$322829.6 enclose $2728$ pixels in total. 
The left panel in Figure~\ref{fig:illustration} shows in red dots 
the $370$ pixels on the boundaries of the feature masks and 
the brightest $2728/2-370=994$ pixels within the feature masks. 
We then map those $1364$ pixels in the image plane back to the source plane 
for a given set of lens parameters, which are the red dots in the right panel. 
The black lines resulting from the Delaunay tessellation connect the $1364$ pixels 
in the source plane. The remaining $1364$ pixels in the image plane will then 
be mapped into individual Delaunay triangles, the weights of which are assumed 
to be inversely proportional to the distances to the three vertices of the 
triangles that they reside in. 

\subsubsection{Optimization}

The parameter optimization is done by minimizing a $\chi^2$ function using 
the Levenberg--Marquardt algorithm as implemented in the {\tt LMFIT} package 
\citep{lmfit}. Each parameter can be assigned a value (required), 
upper and lower bounds (optional), variability (optional), 
and a mathematical expression connecting different parameters (optional). 
For modeling the \Survey{} lenses, the $\chi^2$ function is defined as 
\begin{equation}
\chi^2 = \sum_{i, j} \frac{[I^{\rm data}_{i, j}-(I^{\rm phot}_{i, j}+I^{\rm image}_{i, j})*\text{PSF}]^2}{\sigma_{i, j}^2},
\end{equation}
where the asterisk represents a convolution. 
$I^{\rm data}_{i, j}$, $I^{\rm phot}_{i, j}$, and $I^{\rm image}_{i, j}$ are 
the observed, foreground-light, and model lensed image intensities at pixel 
$(i, j)$ in the image plane, respectively, and $\sigma_{i, j}$ is the corresponding 
rescaled pixel count error. 

Note that {\tt lfit\_gui} also offers another mode for the parameter optimization, 
which is a fully Bayesian method utilizing 
the Affine Invariant Markov Chain Monte Carlo Ensemble sampler \citep{Goodman10} 
implemented in the {\tt emcee} package \citep{emcee}. 
We choose to use the nonlinear Levenberg--Marquardt approach in this work 
because it is sufficient for the purpose of obtaining smooth lens models 
and it significantly reduces the computational time when compared to 
the MCMC approach. 
Nevertheless, the statistical uncertainties estimated from the 
covariance matrix in the nonlinear approach are typically underestimated, and 
the MCMC approach is needed for a better estimate of the parameter uncertainties. 

\begin{table*}[htbp]
\begin{center}
\caption{\label{tb:tb2} Lens parameters of the BELLS GALLERY grade-A lenses.}
\begin{tabular}{c c c c c c c c c c c c}
\hline \hline
Target & \multicolumn{8}{c}{S\'{e}sic} & & \multicolumn{1}{c}{b-spline} & $\mu$ \\ 
\cline{2-9} \cline{11-11}
 & $b_{\rm SIE}$ & $q$ & P.A. & $\Delta$ R.A. & $\Delta$ decl. & $\gamma$ & $\phi_{\gamma}$ & $\chi^2$/dof & & $b_{\rm SIE}$ & \\
 & (arcsec) & & (degree) & (arcsec) & (arcsec) & & (degree) & & & (arcsec) & \\
(1) & (2) & (3) & (4) & (5) & (6) & (7) & (8) & (9) & & (10) & (11) \\
\hline
SDSSJ002927.38$+$254401.7 & 1.34 & 0.717 & \phantom{1}32 & $+0.07$ & $+0.03$ &   &  & 19360/17128 & & 1.34 & 14 \\ 
\hline
SDSSJ011300.57$+$025046.2 & 1.24 & 0.804 & \phantom{1}86 & $-0.08$ & $+0.02$ &   &   & 18133/17120 & & 1.24 & 26 \\ 
 & 0.21 & 0.334 & \phantom{1}37 & $-0.48$ & $+1.68$ & & & & & & \\ 
\hline
SDSSJ020121.39$+$322829.6 & 1.70 & 0.675 & \phantom{1}44 & $-0.04$ & $+0.06$ &   &  & 46375/25895 & & 1.70 & 15 \\ 
\hline
SDSSJ023740.63$-$064112.9 & 0.65 & 0.860 & \phantom{1}49 & $+0.01$ & $+0.05$ &   &  & 9556/10189 & & 0.63 & 9 \\ 
\hline
SDSSJ074249.68$+$334148.9 & 1.22 & 0.741 &  157 & $+0.02$ & $+0.04$ &   &  & 17659/19853 & & 1.22 & 16 \\ 
\hline
SDSSJ075523.52$+$344539.5 & 2.05 & 0.494 &  105 & $+0.07$ & $-0.16$ & 0.24 & 123 & 50182/40380 & & 2.05 & 14 \\ 
\hline
SDSSJ085621.59$+$201040.5 & 0.98 & 0.602 & \phantom{1}90 & $+0.11$ & $+0.05$ &   &  & 13902/14629 & & 0.95 & 6 \\ 
\hline
SDSSJ091807.86$+$451856.7 & 0.77 & 0.889 & \phantom{1}83 & $-0.02$ & $+0.15$ &   &   & 17894/10168 & &   & 7 \\ 
 & 0.11 & 0.132 & \phantom{1}58 & $-0.47$ & $-0.11$ & & &   & & \\ 
\hline
SDSSJ091859.21$+$510452.5 & 1.60 & 0.985 & 157 & $-0.01$ & $+0.02$ & 0.18 & 41 & 43623/25888 & & 1.60 & 18 \\ 
\hline
SDSSJ111027.11$+$280838.4 & 0.98 & 0.917 & \phantom{1}41 & $+0.01$ & $-0.01$ &   &  & 16894/14629 & & 0.98 & 8 \\ 
\hline
SDSSJ111040.42$+$364924.4 & 1.16 & 0.791 & \phantom{1}79 & $+0.01$ & $-0.02$ &   &  & 17709/17135 & & 1.16 & 17 \\ 
\hline
SDSSJ111634.55$+$091503.0 & 1.03 & 0.850 & \phantom{11}0 & $-0.02$ & $+0.00 $ &   &  & 20264/19871 & & 1.05 & 4 \\ 
\hline
SDSSJ114154.71$+$221628.8 & 1.27 & 0.768 &  154 & $+0.05$ & $+0.02$ &   &  & 21172/19862 & & 1.27 & 8 \\ 
\hline
SDSSJ120159.02$+$474323.2 & 1.18 & 0.810 & \phantom{1}39 & $-0.06$ & $+0.01$ &   &  & 17580/14608 & & 1.18 & 12 \\ 
\hline
SDSSJ122656.45$+$545739.0 & 1.37 & 0.923 & \phantom{1}37 & $+0.02$ & $+0.00 $ & 0.15 & 67 & 23584/22782 & & 1.38 & 13 \\ 
\hline
SDSSJ222825.76$+$120503.9 & 1.28 & 0.792 &  177 & $+0.02$ & $+0.08$ &   &  & 14788/10168 & & 1.31 & 6 \\ 
\hline
SDSSJ234248.68$-$012032.5 & 1.11 & 0.827 & \phantom{1}53 & $-0.04$ & $+0.01$ &   &  & 16168/14615 & & 1.11 & 23 \\ 
\hline \hline
\end{tabular}
\end{center}
\textsc{      Note.} --- Column 1 is the SDSS system name. Columns 2-8 are the Einstein radius, minor-to-major axis ratio, major-axis position angle with respect to the north, centroid offset in R.A. and decl. of the SIE component, and strength and polar angle of the external shear found using the S\'{e}rsic model for foreground-light subtraction. Column 9 provides the $\chi^2$ value and the number of degrees of freedom (dof). Column 10 is the Einstein radius found using the b-spline model for foreground-light subtraction. Column 11 is the average magnification found using the S\'{e}rsic model for foreground-light subtraction. For SDSSJ011300.57$+$025046.2 and SDSSJ091807.86$+$451856.7, the second row provides the parameters of the second SIE component. \\
\end{table*}
\begin{table*}[htbp]
\begin{center}
\caption{\label{tb:tb3} Notes on Five Grade-A lenses with special treatments.}
\begin{tabular}{c l}
\hline \hline
Target & Comments \\
\hline
SDSSJ011300.57$+$025046.2 & The arc near the location of the bright galaxy to 
the north of the lens galaxy is bent slightly inward. \\ 
& A second SIE mass component is used to model 
this perturber. Although another bright galaxy is seen \\
& to the south of the lens galaxy, the inclusion of 
a third SIE component does not yield a significantly better fit. \\
SDSSJ075523.52$+$344539.5 & A quadruple-image plus a double-image system. Several 
nearby luminous clumps are clearly seen. External \\
& shear is needed to correctly model the positions of the quadruple images.  The inferred mass center is \\ 
& offset from the light center by 0\farcs17 which is suspected to be the result of the local environment. \\
& Fixing the mass center to the light center yields a significantly worse model with $\Delta \chi^2 \sim 40,000$. \\
SDSSJ091807.86$+$451856.7 & The foreground lens is clearly composed of two distinct 
components. A two-SIE lens model is needed, \\
& though the second mass component has an unphysically small axis ratio 
in the best-fit model. The mass \\
&centers are offset from the light centers by 0\farcs13 and 0\farcs12, respectively. Fixing the mass \\
& centers to the light centers yields a slightly worse model with $\Delta \chi^2 \sim 500$. The system is \\
& rather complicated, and a detailed investigation of the mass/light offset is needed. \\
SDSSJ091859.21$+$510452.5 & A quadruple-image system in a cusp configuration. A 
perturber is seen to the west of the main lens galaxy, \\
& which seems to further reside in a crowded environment.
External shear is needed. \\
SDSSJ122656.45$+$545739.0 & A very beautiful and rare system with two sets of 
quadruple images. Several luminous clumps are seen \\ 
& within 10$^{\prime \prime}$ distance.  
External shear is needed to correctly reproduce the positions of all the eight images. \\
\hline \hline
\end{tabular}
\end{center}
\end{table*}

\subsection{Results}
\label{sect:results}

Table~\ref{tb:tb2} provides the lens model parameters of the 17 \Survey{} grade-A 
lenses obtained from this nonlinear optimization approach. 
Thirteen systems can be well explained by a simple SIE lens model. 
Substantial amounts of external shear are also required for J075523.52$+$344539.5, 
J091859.21$+$510452.5, and J122656.45$+$545739.0 
which are also seen to reside in crowded environments. 
J011300.57$+$025046.2 and J091807.86$+$451856.7 each 
require the inclusion of a second SIE mass component 
to model the gravitational effects from nearby luminous perturbers. 
Detailed descriptions of these five systems are given in Table~\ref{tb:tb3}. 

\begin{table*}[htbp]
\begin{center}
\caption{\label{tb:tb_source} Source parameters of the BELLS GALLERY grade-A lenses.}
\begin{tabular}{c c c c c c c c c c}
\hline \hline
Target & Source & $\Delta$R.A. & $\Delta$decl. & $q$ & $n$  & $R_{\rm eff}$ & $R_{\rm eff}$ & $\Delta$$d$ & $m_{\rm AB}$ \\ 
 & ID & (arcsec) & (arcsec) & & & (arcsec) & (pc) & (pc) & (mag) \\
(1) & (2) & (3) & (4) & (5) & (6) & (7) & (8) & (9) & (10) \\
\hline
SDSSJ002927.38$+$254401.7 & S1 & $ -0.08 $ &$ -0.10 $ &$ 0.08$ & $0.33$ & $ 0.0062 \pm 0.0011 $ & $51 \pm 8 $ & 0 & $ 25.3 \pm 1.2 $ \\
 & S2 & $ -0.16 $ &$ +0.12 $ &$ 0.52$ & $4.02$ & $ 0.2931 \pm 0.1163 $ & $2435 \pm 966 $ & 1944 & $ 25.6 \pm 1.4 $ \\
 & S3 & $ -0.07 $ &$ -0.11 $ &$ 0.59$ & $1.57$ & $ 0.0173 \pm 0.0011 $ & $144 \pm 8 $ & 94 & $ 26.1 \pm 0.6 $ \\
 & S4 & $ -0.13 $ &$ -0.09 $ &$ 0.15$ & $3.00$ & $ 0.0640 \pm 0.0125 $ & $531 \pm 103 $ & 413 & $ 27.3 \pm 0.9 $ \\
\hline
SDSSJ011300.57$+$025046.2 & S1 & $ -0.03 $ &$ +0.19 $ &$ 0.57$ & $3.10$ & $ 0.1267 \pm 0.0341 $ & $1039 \pm 279 $ & 0 & $ 27.1 \pm 0.7 $ \\
 & S2 & $ -0.27 $ &$ +0.15 $ &$ 0.13$ & $1.40$ & $ 0.0543 \pm 0.0041 $ & $445 \pm 33 $ & 2002 & $ 27.1 \pm 0.3 $ \\
\hline
SDSSJ020121.39$+$322829.6 & S1 & $ +0.25 $ &$ -0.25 $ &$ 0.17$ & $4.66$ & $ 0.1326 \pm 0.0112 $ & $1066 \pm 89 $ & 0 & $ 24.8 \pm 0.2 $ \\
 & S2 & $ +0.24 $ &$ -0.10 $ &$ 0.40$ & $3.44$ & $ 0.2183 \pm 0.0277 $ & $1756 \pm 223 $ & 1202 & $ 24.5 \pm 0.5 $ \\
 & S3 & $ +0.21 $ &$ -0.30 $ &$ 0.72$ & $0.18$ & $ 0.1020 \pm 0.0007 $ & $821 \pm 5 $ & 558 & $ 24.8 \pm 0.1 $ \\
\hline
SDSSJ023740.63$-$064112.9 & S1 & $ -0.08 $ &$ +0.01 $ &$ 0.46$ & $1.46$ & $ 0.0206 \pm 0.0023 $ & $174 \pm 19 $ & 0 & $ 27.4 \pm 0.4 $ \\
\hline
SDSSJ074249.68$+$334148.9 & S1 & $ -0.22 $ &$ +0.15 $ &$ 0.17$ & $1.68$ & $ 0.0138 \pm 0.0007 $ & $115 \pm 6 $ & 0 & $ 27.1 \pm 0.5 $ \\
 & S2 & $ -0.17 $ &$ +0.10 $ &$ 0.30$ & $0.05$ & $ 0.0194 \pm 0.5311 $ & $162 \pm 4444 $ & 658 & $ 27.3 \pm 0.1 $ \\
 & S3 & $ -0.23 $ &$ +0.13 $ &$ 0.65$ & $2.28$ & $ 0.1172 \pm 0.0077 $ & $980 \pm 64 $ & 179 & $ 25.6 \pm 0.2 $ \\
\hline
SDSSJ075523.52$+$344539.5 & S1 & $ +0.07 $ &$ -0.36 $ &$ 0.52$ & $1.07$ & $ 0.1294 \pm 0.0013 $ & $1058 \pm 10 $ & 0 & $ 24.5 \pm 0.1 $ \\
 & S2 & $ -0.01 $ &$ -0.28 $ &$ 0.47$ & $0.33$ & $ 0.0324 \pm 0.0003 $ & $264 \pm 2 $ & 999 & $ 25.5 \pm 0.1 $ \\
 & S3 & $ +0.65 $ &$ +0.48 $ &$ 0.42$ & $0.57$ & $ 0.1431 \pm 0.0062 $ & $1170 \pm 50 $ & 8339 & $ 26.1 \pm 0.2 $ \\
\hline
SDSSJ085621.59$+$201040.5 & S1 & $ -0.34 $ &$ +0.16 $ &$ 0.60$ & $5.39$ & $ 0.0617 \pm 0.0090 $ & $521 \pm 76 $ & 0 & $ 25.5 \pm 2.3 $ \\
\hline
SDSSJ091807.86$+$451856.7 & S1 & $ +0.06 $ &$ -0.02 $ &$ 0.66$ & $2.32$ & $ 0.0209 \pm 0.0009 $ & $175 \pm 7 $ & 0 & $ 25.6 \pm 0.5 $ \\
 & S2 & $ +0.27 $ &$ +0.01 $ &$ 0.37$ & $1.78$ & $ 0.3800 \pm 0.0468 $ & $3184 \pm 392 $ & 1793 & $ 24.5 \pm 0.2 $ \\
\hline
SDSSJ091859.21$+$510452.5 & S1 & $ +0.08 $ &$ -0.27 $ &$ 0.09$ & $0.06$ & $ 0.0103 \pm 0.0042 $ & $86 \pm 35 $ & 0 & $ 24.9 \pm 0.1 $ \\
 & S2 & $ +0.27 $ &$ -0.15 $ &$ 0.29$ & $0.19$ & $ 0.0109 \pm 0.0006 $ & $90 \pm 5 $ & 1909 & $ 27.8 \pm 0.1 $ \\
 & S3 & $ +0.17 $ &$ -0.22 $ &$ 0.34$ & $1.73$ & $ 0.1478 \pm 0.0044 $ & $1232 \pm 36 $ & 894 & $ 24.7 \pm 0.1 $ \\
\hline
SDSSJ111027.11$+$280838.4 & S1 & $ +0.18 $ &$ +0.21 $ &$ 0.24$ & $1.27$ & $ 0.0436 \pm 0.0022 $ & $363 \pm 18 $ & 0 & $ 26.2 \pm 0.2 $ \\
\hline
SDSSJ111040.42$+$364924.4 & S1 & $ -0.02 $ &$ -0.07 $ &$ 0.65$ & $0.52$ & $ 0.0180 \pm 0.0004 $ & $148 \pm 3 $ & 0 & $ 27.0 \pm 0.1 $ \\
 & S2 & $ -0.03 $ &$ -0.23 $ &$ 0.44$ & $0.54$ & $ 0.0495 \pm 0.0013 $ & $410 \pm 10 $ & 1278 & $ 26.6 \pm 0.1 $ \\
 & S3 & $ +0.00 $ &$ -0.01 $ &$ 0.75$ & $0.42$ & $ 0.0243 \pm 0.0006 $ & $201 \pm 5 $ & 582 & $ 27.6 \pm 0.1 $ \\
 & S4 & $ +0.03 $ &$ -0.10 $ &$ 0.87$ & $2.29$ & $ 0.3174 \pm 0.0495 $ & $2626 \pm 409 $ & 421 & $ 25.2 \pm 0.2 $ \\
\hline
SDSSJ111634.55$+$091503.0 & S1 & $ +0.05 $ &$ -0.61 $ &$ 0.17$ & $0.05$ & $ 0.0259 \pm 0.1889 $ & $215 \pm 1569 $ & 0 & $ 25.1 \pm 0.1 $ \\
 & S2 & $ -0.01 $ &$ -0.55 $ &$ 0.64$ & $0.90$ & $ 0.1008 \pm 0.0057 $ & $837 \pm 47 $ & 748 & $ 25.5 \pm 0.1 $ \\
\hline
SDSSJ114154.71$+$221628.8 & S1 & $ -0.25 $ &$ -0.03 $ &$ 0.79$ & $4.72$ & $ 0.0298 \pm 0.0018 $ & $240 \pm 14 $ & 0 & $ 24.9 \pm 1.2 $ \\
 & S2 & $ -0.21 $ &$ +0.07 $ &$ 0.49$ & $0.05$ & $ 0.0195 \pm 0.1626 $ & $157 \pm 1315 $ & 882 & $ 27.8 \pm 0.1 $ \\
\hline
SDSSJ120159.02$+$474323.2 & S1 & $ +0.02 $ &$ +0.11 $ &$ 0.79$ & $4.47$ & $ 0.2796 \pm 0.0432 $ & $2378 \pm 367 $ & 0 & $ 24.2 \pm 0.6 $ \\
 & S2 & $ +0.11 $ &$ +0.03 $ &$ 0.46$ & $0.05$ & $ 0.0269 \pm 1.9713 $ & $228 \pm 16768 $ & 1038 & $ 26.6 \pm 0.1 $ \\
 & S3 & $ -0.23 $ &$ +0.28 $ &$ 0.56$ & $1.31$ & $ 0.0970 \pm 0.0089 $ & $825 \pm 75 $ & 2554 & $ 26.5 \pm 0.2 $ \\
 & S4 & $ +0.03 $ &$ +0.09 $ &$ 0.04$ & $0.28$ & $ 0.0037 \pm 0.0002 $ & $31 \pm 1 $ & 185 & $ 29.6 \pm 0.2 $ \\
\hline
SDSSJ122656.45$+$545739.0 & S1 & $ -0.09 $ &$ -0.10 $ &$ 0.41$ & $13.21$ & $ 2.7357 \pm 5.3906 $ & $22190 \pm 43725 $ & 0 & $ 24.7 \pm 9.1 $ \\
 & S2 & $ +0.18 $ &$ +0.00 $ &$ 0.67$ & $5.07$ & $ 0.0135 \pm 0.0021 $ & $109 \pm 17 $ & 2358 & $ 26.9 \pm 3.6 $ \\
 & S3 & $ -0.17 $ &$ +0.01 $ &$ 0.37$ & $0.48$ & $ 0.0447 \pm 0.0019 $ & $362 \pm 15 $ & 1181 & $ 27.8 \pm 0.2 $ \\
 & S4 & $ -0.05 $ &$ -0.13 $ &$ 0.19$ & $0.36$ & $ 0.0033 \pm 0.0339 $ & $27 \pm 275 $ & 353 & $ 21.2 \pm 123.4 $ \\
\hline
SDSSJ222825.76$+$120503.9 & S1 & $ +0.26 $ &$ -0.13 $ &$ 0.41$ & $1.80$ & $ 0.0077 \pm 0.0045 $ & $61 \pm 35 $ & 0 & $ 25.5 \pm 3.7 $ \\
 & S2 & $ +0.30 $ &$ -0.25 $ &$ 0.61$ & $3.13$ & $ 0.2470 \pm 0.0627 $ & $1985 \pm 504 $ & 956 & $ 25.5 \pm 0.9 $ \\
\hline
SDSSJ234248.68$-$012032.5 & S1 & $ -0.07 $ &$ -0.09 $ &$ 0.62$ & $2.26$ & $ 0.0998 \pm 0.0117 $ & $841 \pm 98 $ & 0 & $ 26.4 \pm 0.4 $ \\
 & S2 & $ -0.04 $ &$ -0.05 $ &$ 0.49$ & $0.05$ & $ 0.0058 \pm 0.0321 $ & $49 \pm 270 $ & 465 & $ 28.5 \pm 0.1 $ \\
 & S3 & $ -0.15 $ &$ -0.23 $ &$ 0.50$ & $0.05$ & $ 0.0198 \pm 0.4625 $ & $167 \pm 3898 $ & 1302 & $ 28.5 \pm 0.2 $ \\
\hline \hline
\end{tabular}
\end{center}
\textsc{      Note.} --- Column 1 is the SDSS system name. Columns 2-8 provide ID, central positions relative to the center of the cutout in R.A. and decl., minor-to-major axis ratio, S\'{e}rsic index, effective radius in arcsec, and effective radius in parsecs of each individual source component. Column 9 is the projected separation from the first source component in parsecs. Column 10 is the magnification-corrected rest-frame far UV apparent AB magnitude of each individual source component calculated from the best-fit source model. \\
\end{table*}

The best-fit lens parameters for the two foreground-light 
subtraction schemes are almost identical for 13 of the 17 grade-A lenses. 
The four systems with significantly larger relative deviations in the 
Einstein radius 
are J023740.63-064112.9 ($3.0\%$), J085621.59$+$201040.5 ($3.0\%$), 
J111634.55$+$091503.0 ($1.8\%$), and J222825.76$+$120503.9 ($1.9\%$).
These four systems turn out to have relatively smaller Einstein radii and 
similar two image configurations with one image very close to the lens 
galaxy. This likely makes the results more sensitive to the method used to subtract 
the unlensed light. 
Consequently, we only report results using the joint modeling approach. 
The center of the SIE component is allowed to vary during the lens modeling. 
In most cases, the inferred lens position is consistent with the observed position 
of the lens galaxies given the typical model position uncertainties of 0\farcs04 
(see Columns 5 and 6 of Table~\ref{tb:tb2}). Two 
systems, J075523.52$+$344539.5 and J091807.86$+$451856.7, 
show significant spatial offsets ($> 0\farcs12$) between the mass and light 
and are discussed further in Table~\ref{tb:tb3}.
Because the source and lensed images are spatially extended, we define the 
average magnification to be the ratio of the observed total light 
of the lensed images to that of the source. 
The background LAEs are magnified by factors from 
$4$ to $26$ with a median magnification of $13$. 

\begin{figure*}[htbp]
\centering
\includegraphics[width=0.98\textwidth]{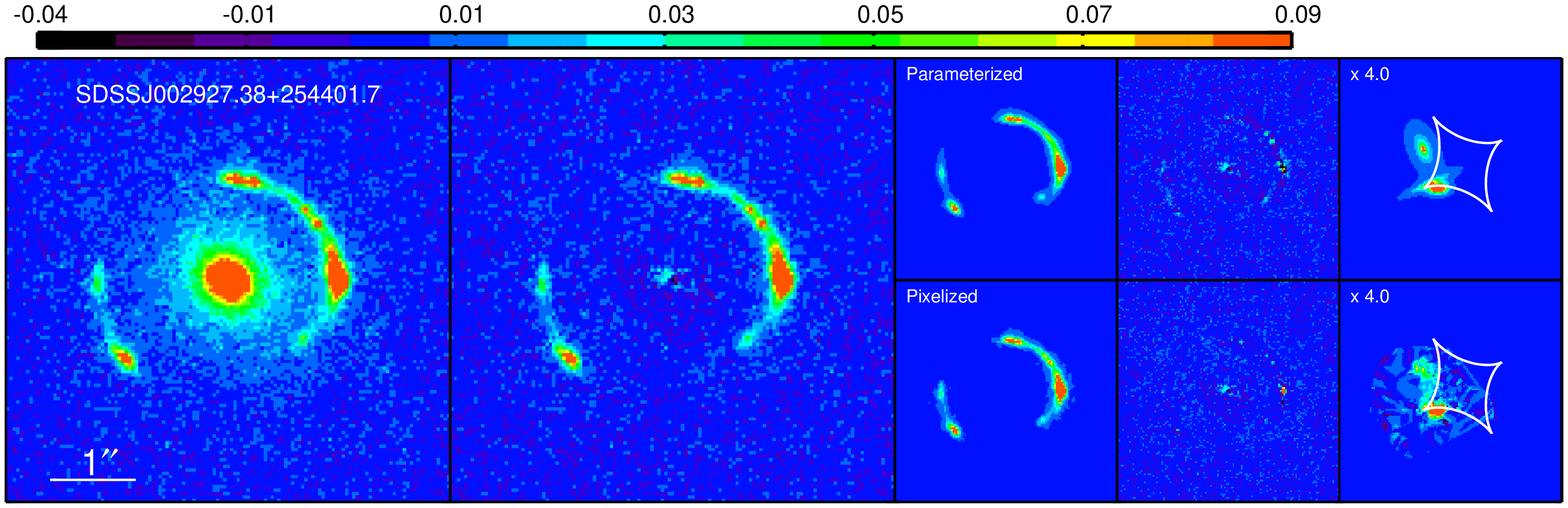}
\includegraphics[width=0.98\textwidth]{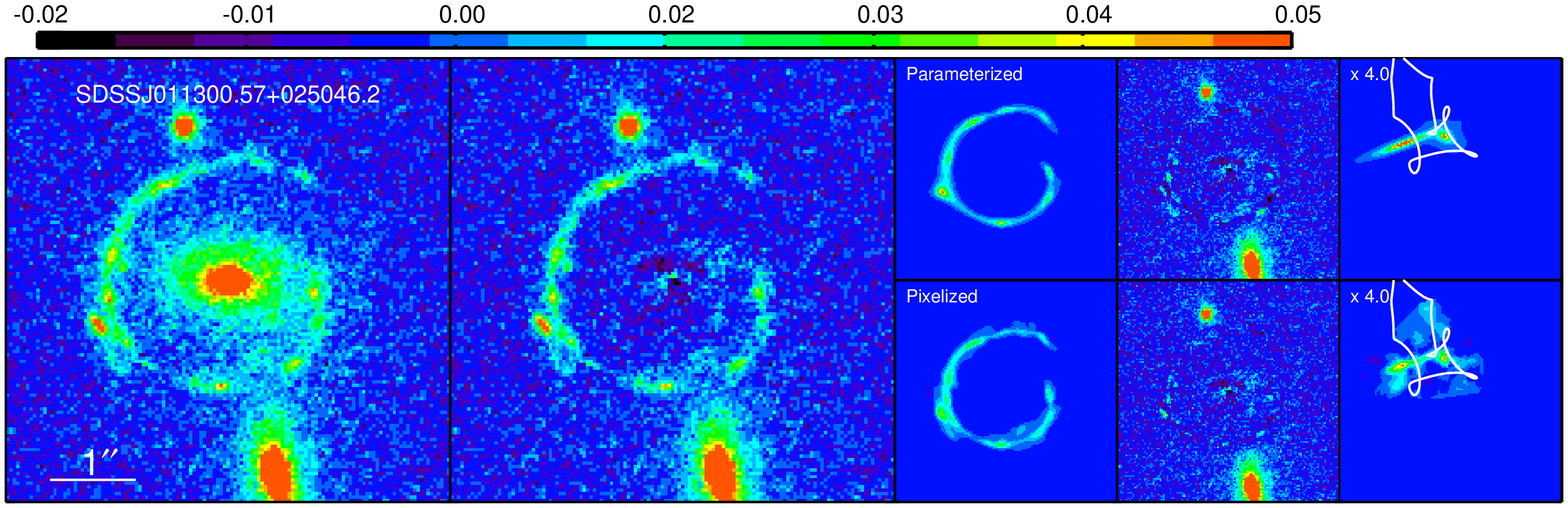}
\includegraphics[width=0.98\textwidth]{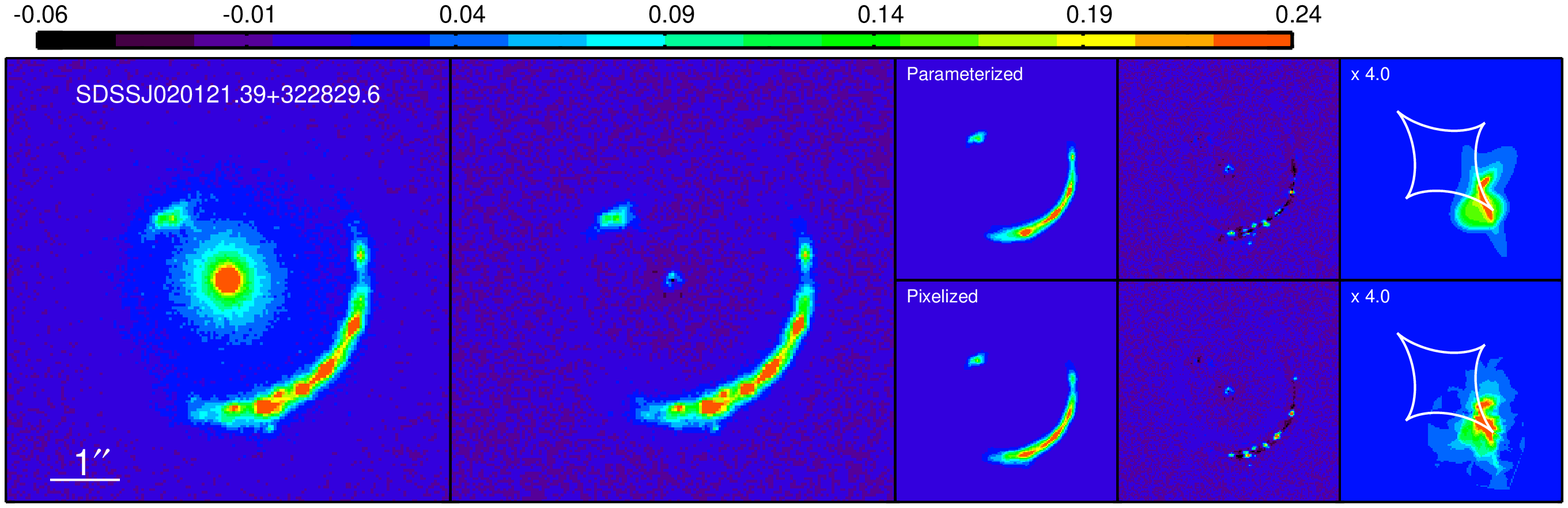}
\includegraphics[width=0.98\textwidth]{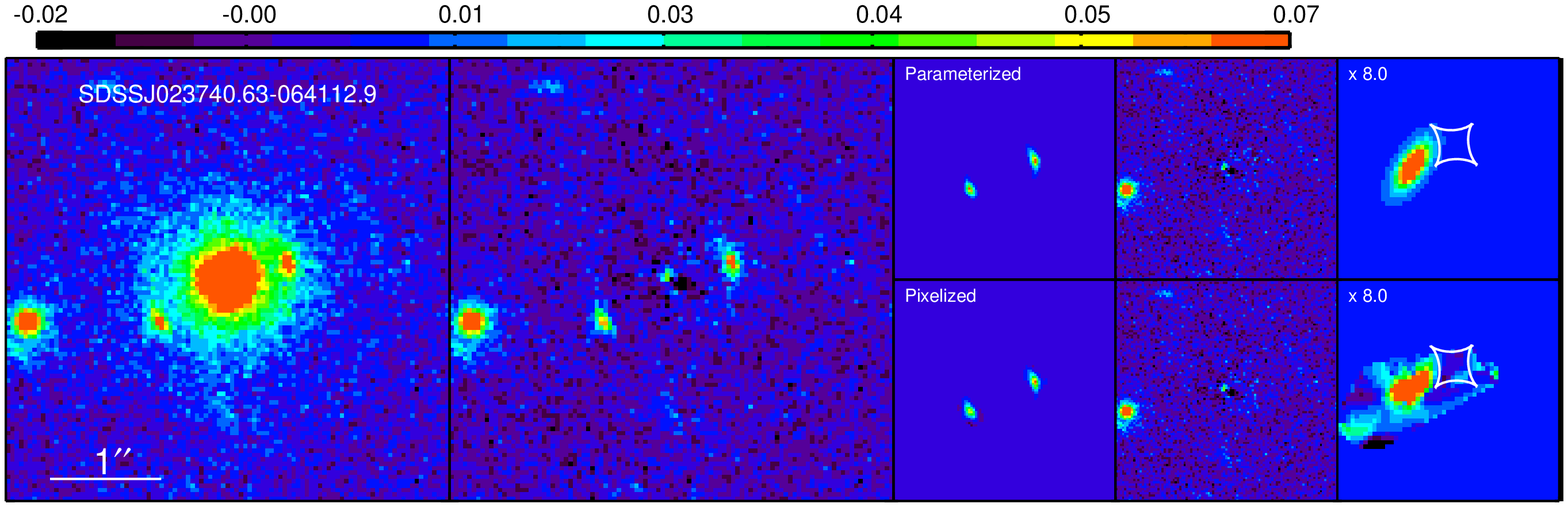}
\caption{\label{fig:models_1}
Smooth lens models for the 17 \Survey{} grade-A lenses. 
The observational data, foreground-subtracted image, predicted lensed image, 
final residual, and the background source model are shown from left to right, 
respectively. For each system, the results of the two source models are 
split into two rows with the parameterized source model on the top and 
the pixelized source model on the bottom. The white lines in the last 
panels are the caustics of the lens model. All the images are orientated such that 
north is up and east is to the left. The source plane panels are magnified by 
factors of four or eight relative to the image plane panel as indicated in 
each panel. The color bars indicate the intensity levels in units of 
electrons per second per pixel$^2$. }
\end{figure*}
\addtocounter{figure}{-1}
\begin{figure*}[htbp]
\centering
\includegraphics[width=0.98\textwidth]{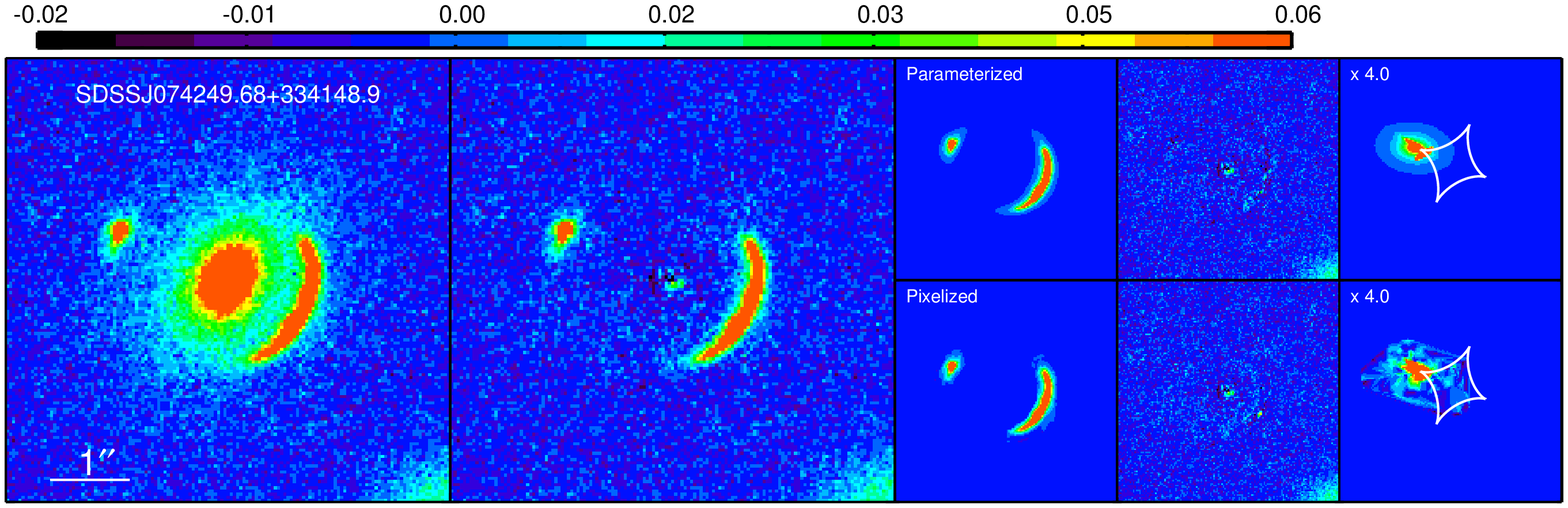}
\includegraphics[width=0.98\textwidth]{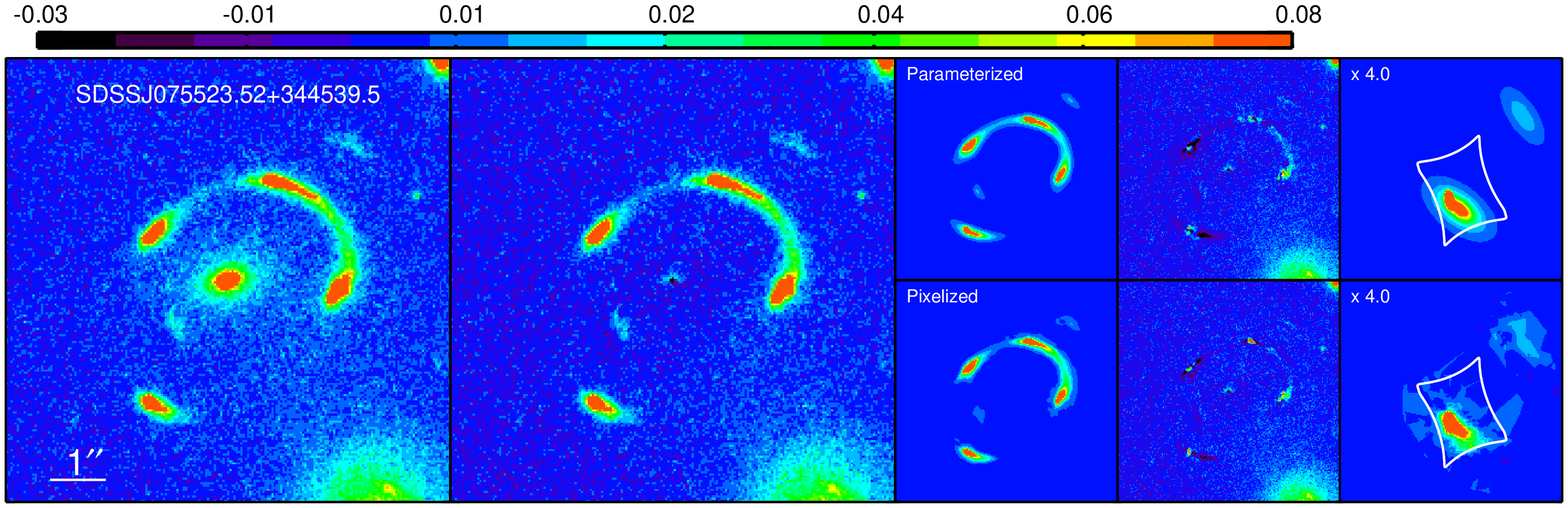}
\includegraphics[width=0.98\textwidth]{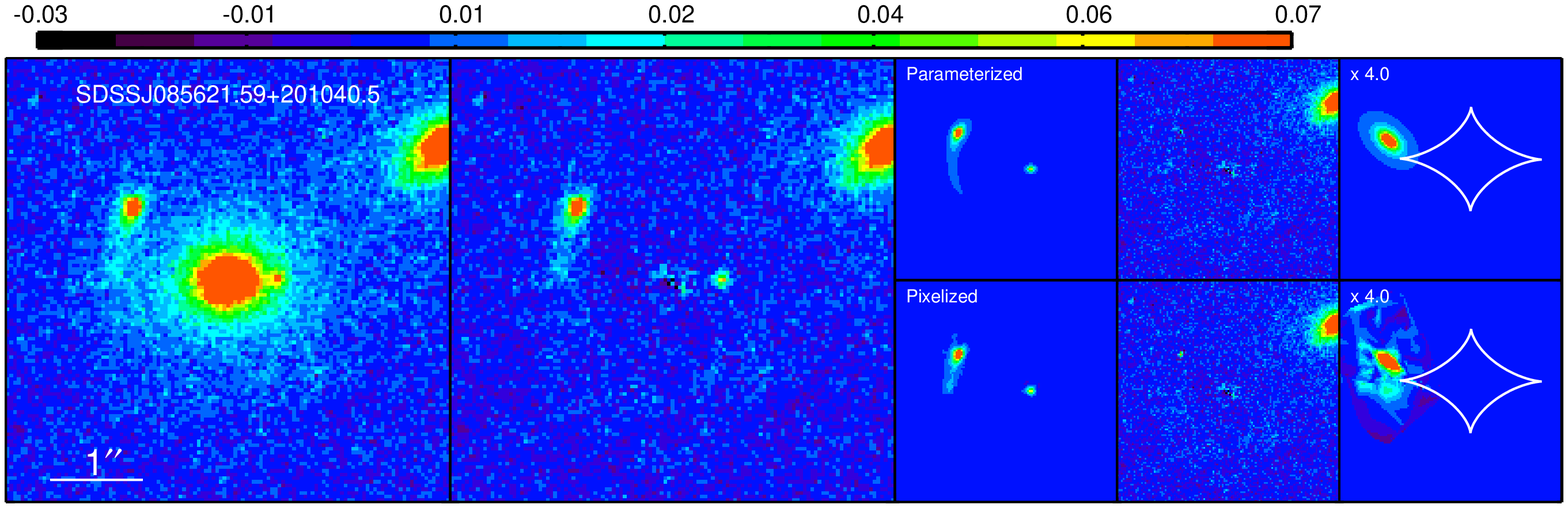}
\includegraphics[width=0.98\textwidth]{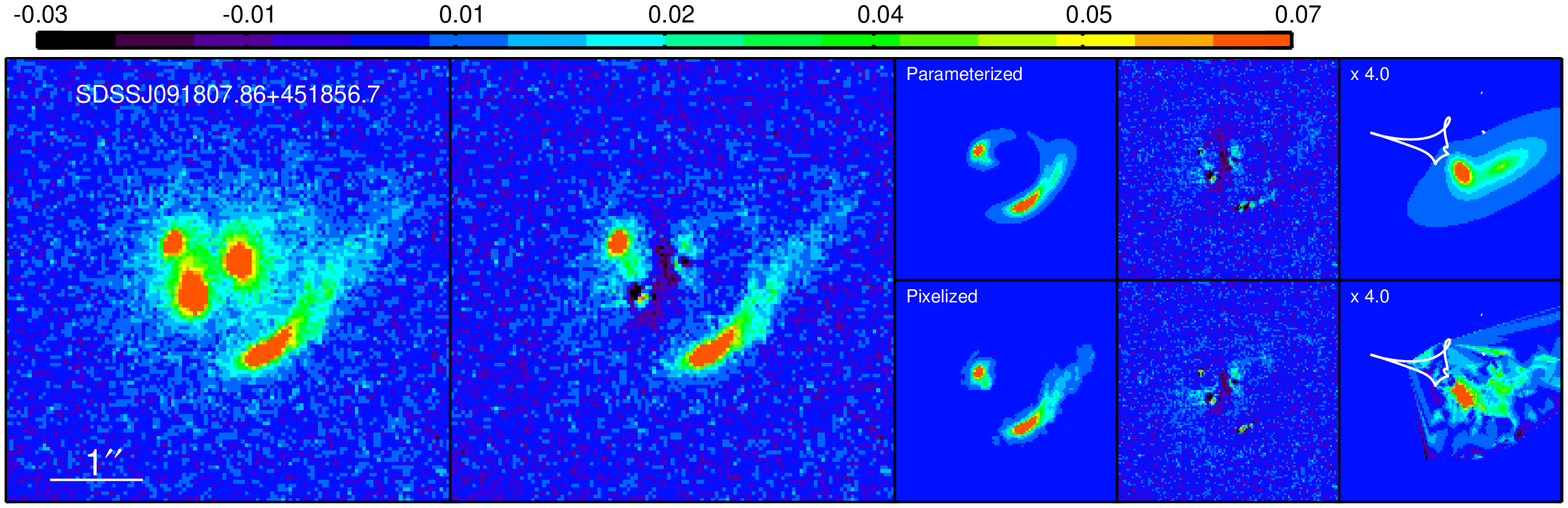}
\caption{{\label{fig:models_2}} \textit{Continued}}
\end{figure*}
\addtocounter{figure}{-1}
\begin{figure*}[htbp]
\centering
\includegraphics[width=0.98\textwidth]{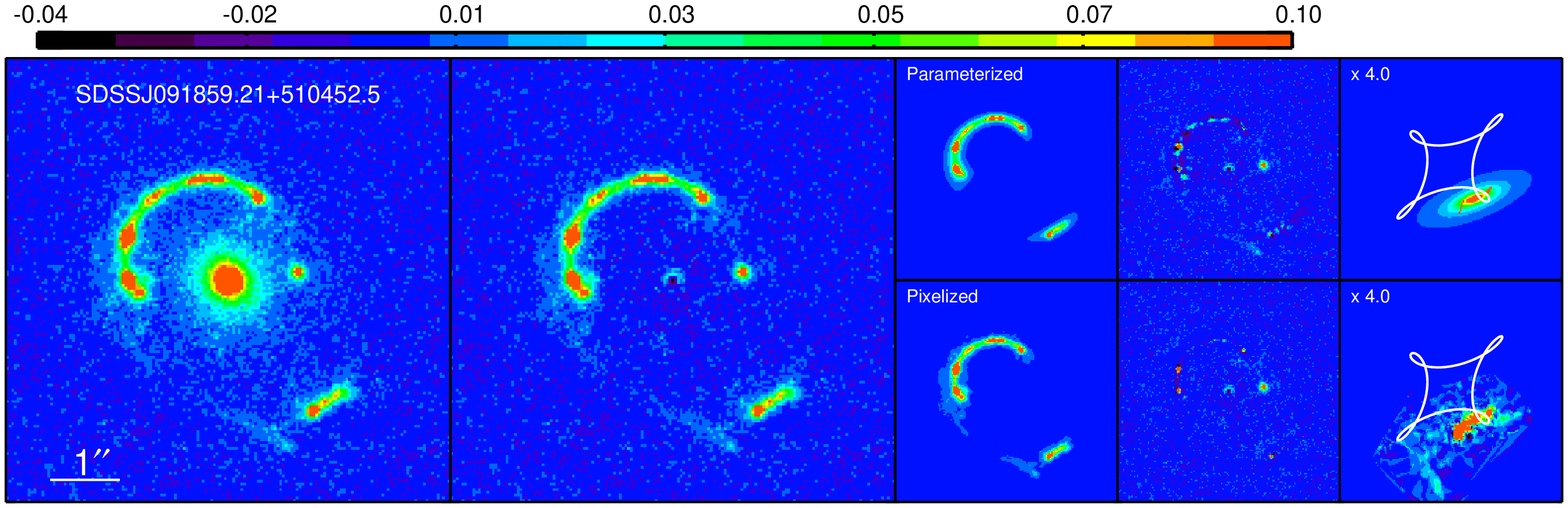}
\includegraphics[width=0.98\textwidth]{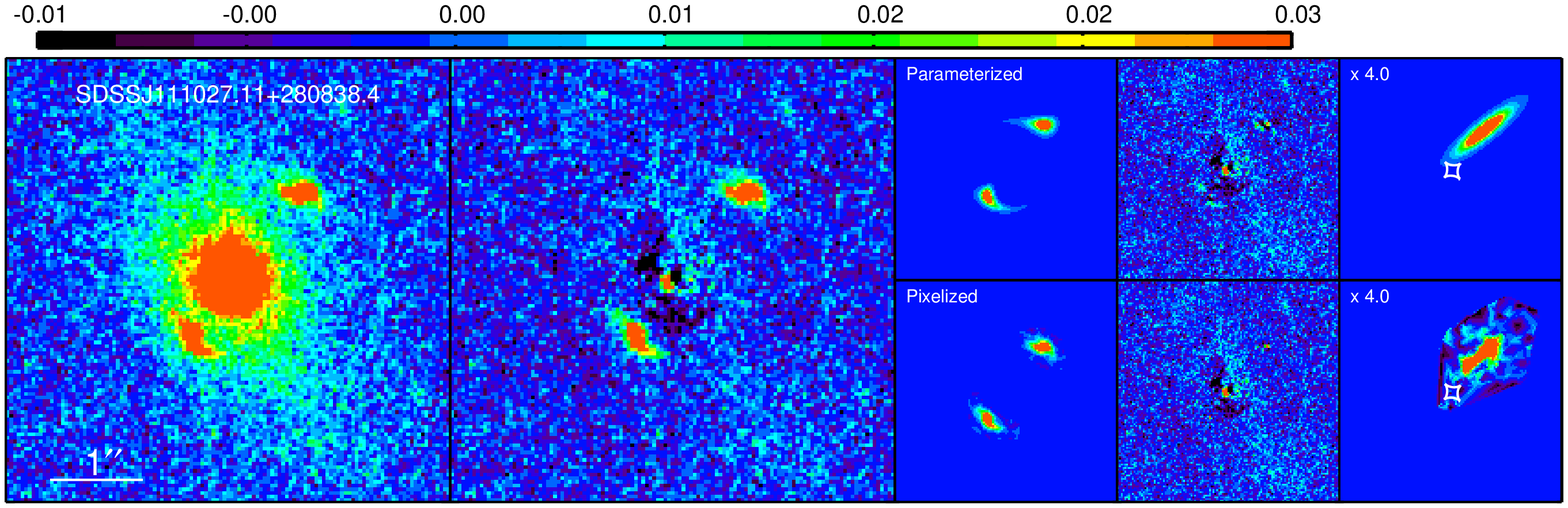}
\includegraphics[width=0.98\textwidth]{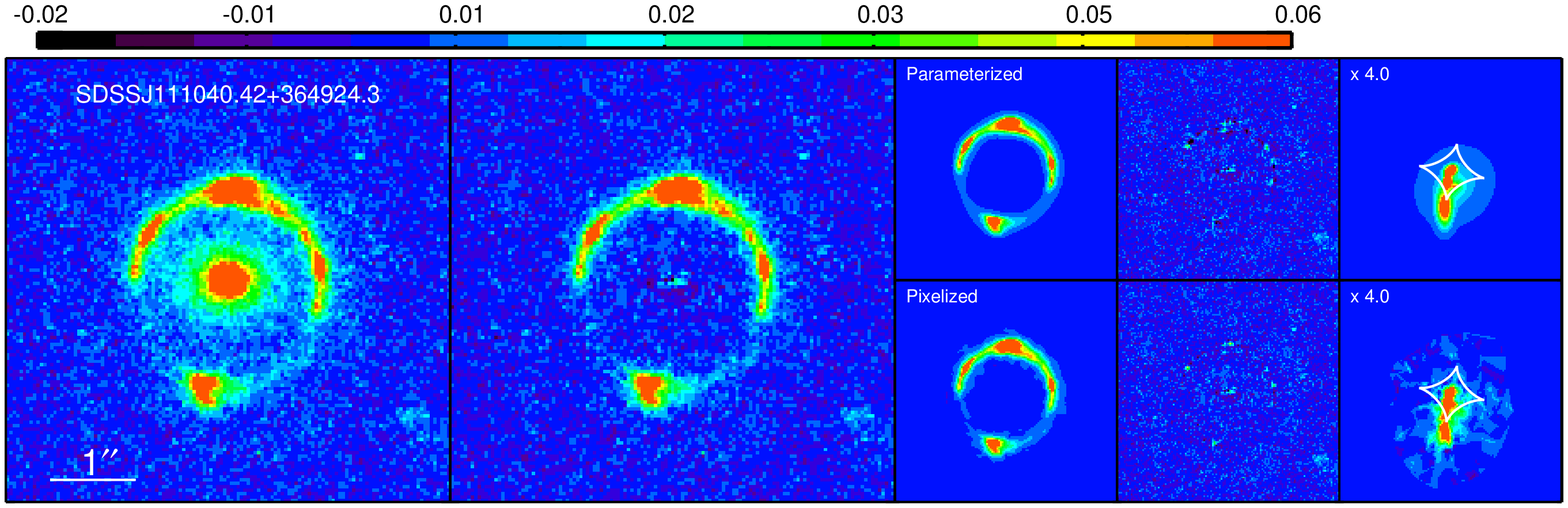}
\includegraphics[width=0.98\textwidth]{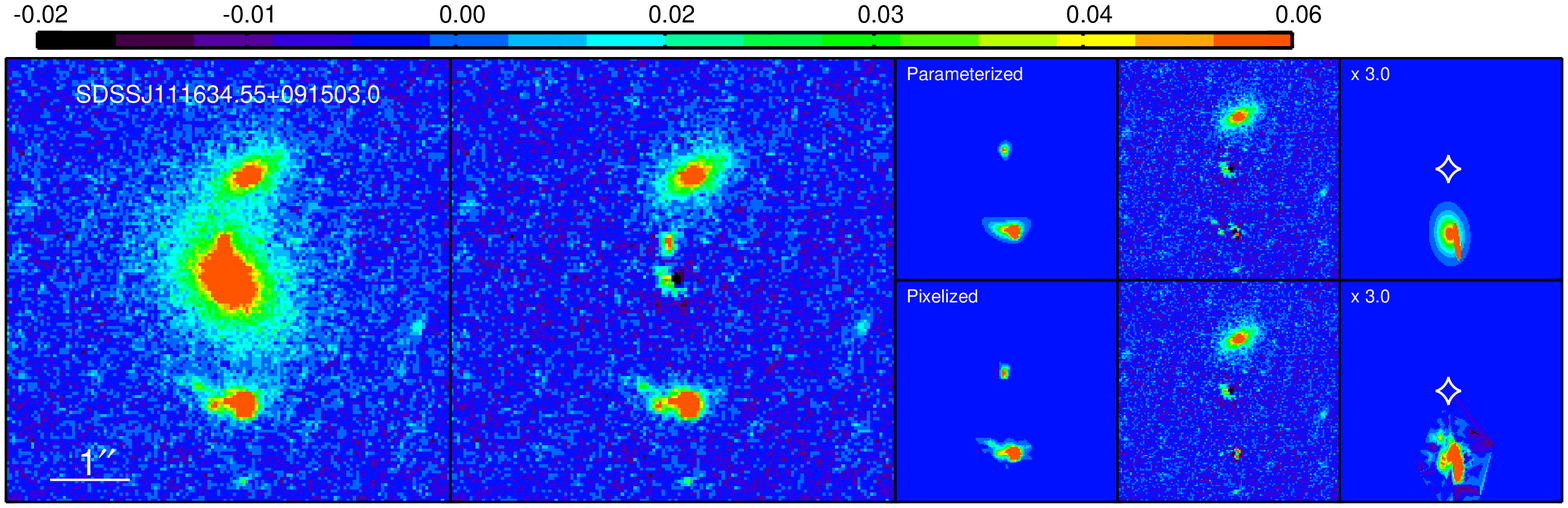}
\caption{{\label{fig:models_3}} \textit{Continued}}
\end{figure*}
\addtocounter{figure}{-1}
\begin{figure*}[htbp]
\centering
\includegraphics[width=0.98\textwidth]{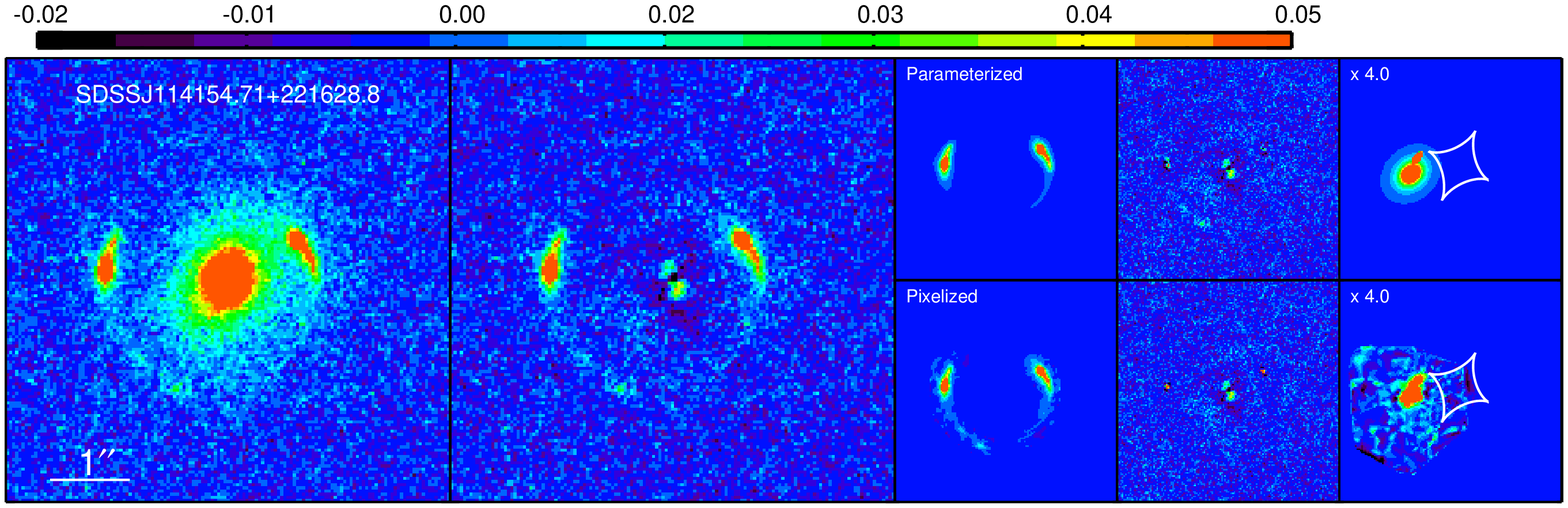}
\includegraphics[width=0.98\textwidth]{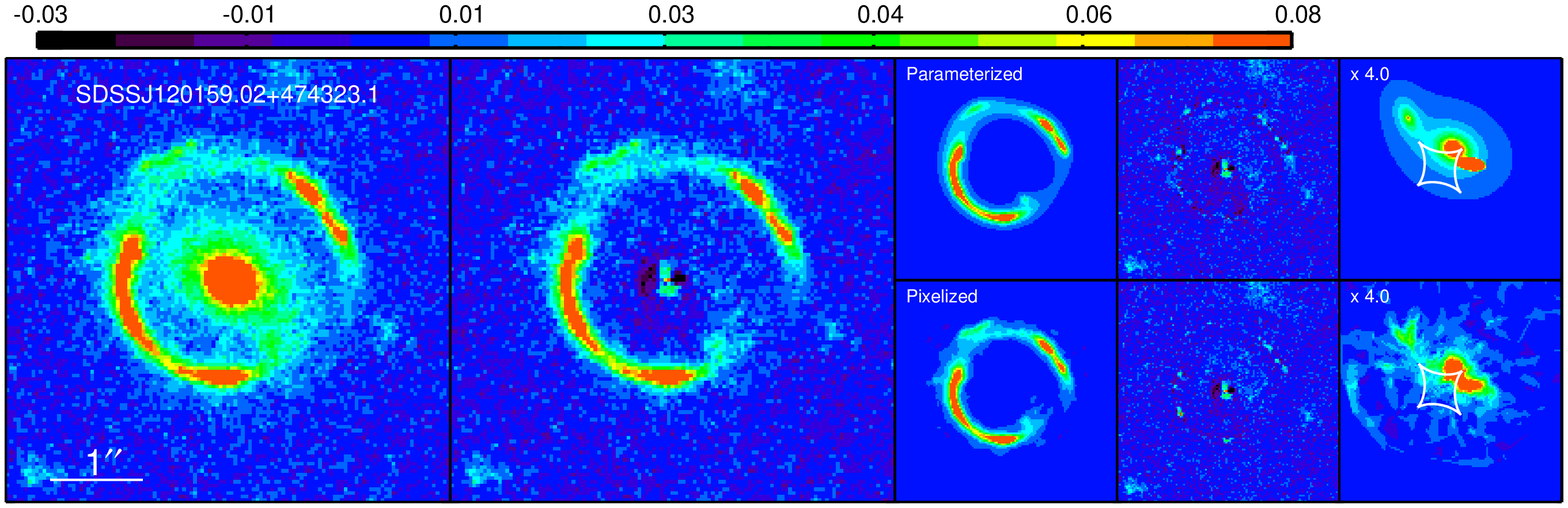}
\includegraphics[width=0.98\textwidth]{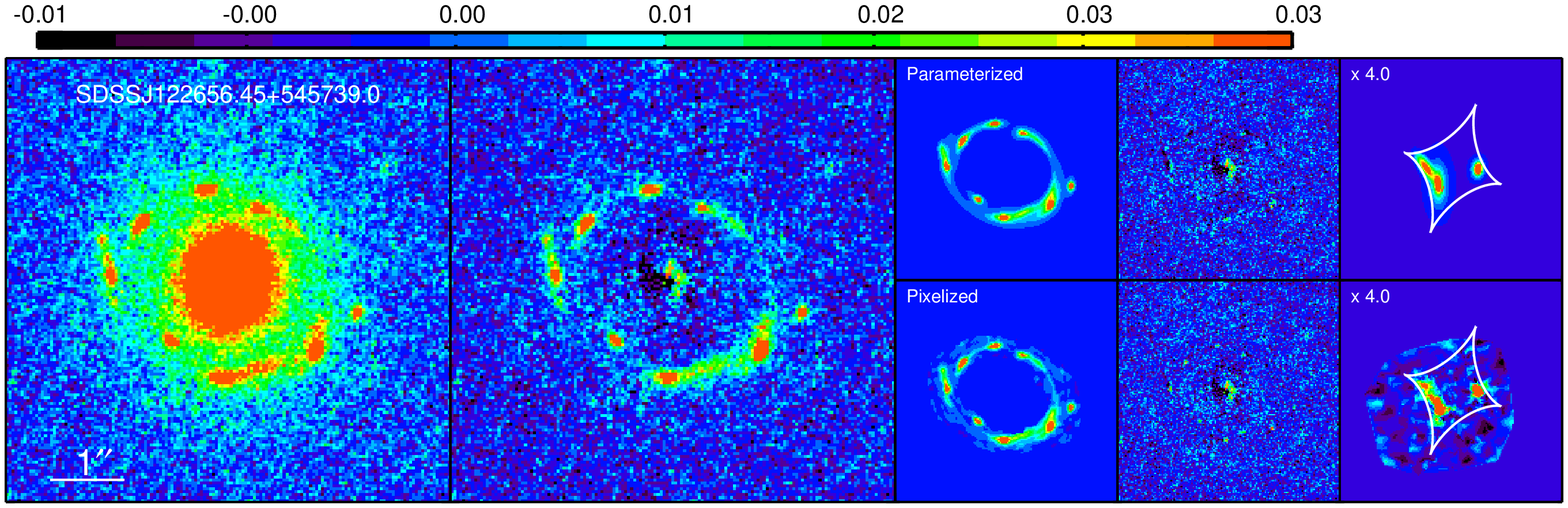}
\includegraphics[width=0.98\textwidth]{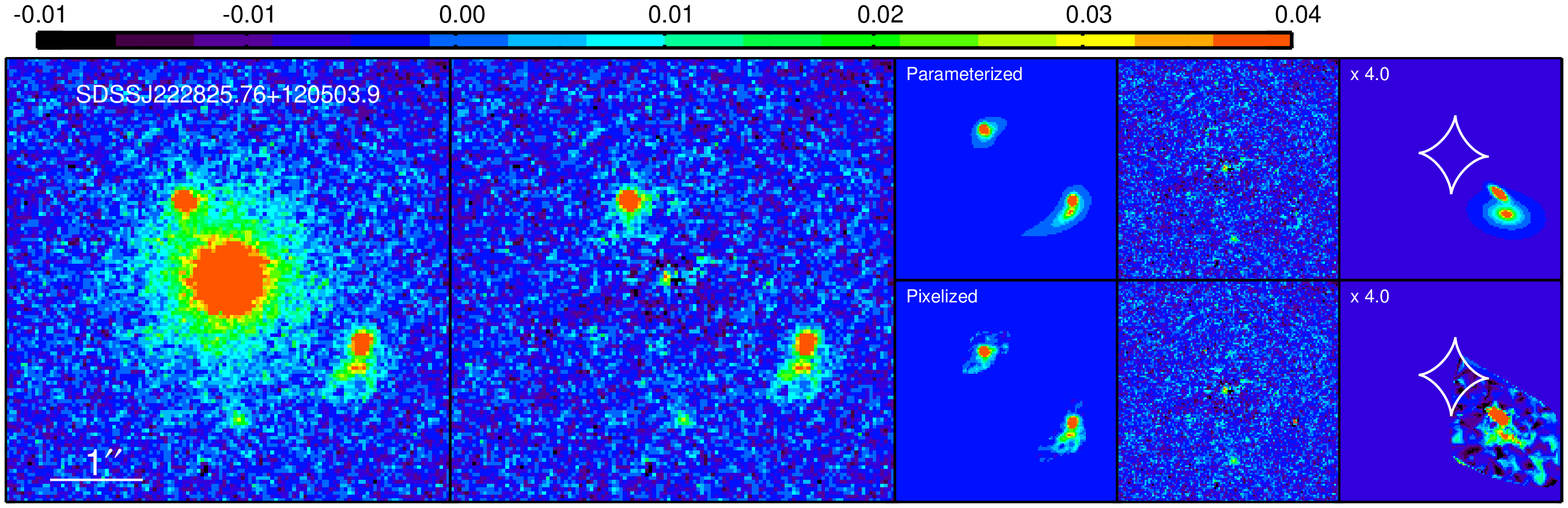}
\caption{{\label{fig:models_4}} \textit{Continued}}
\end{figure*}
\addtocounter{figure}{-1}
\begin{figure*}[htbp]
\centering
\includegraphics[width=0.98\textwidth]{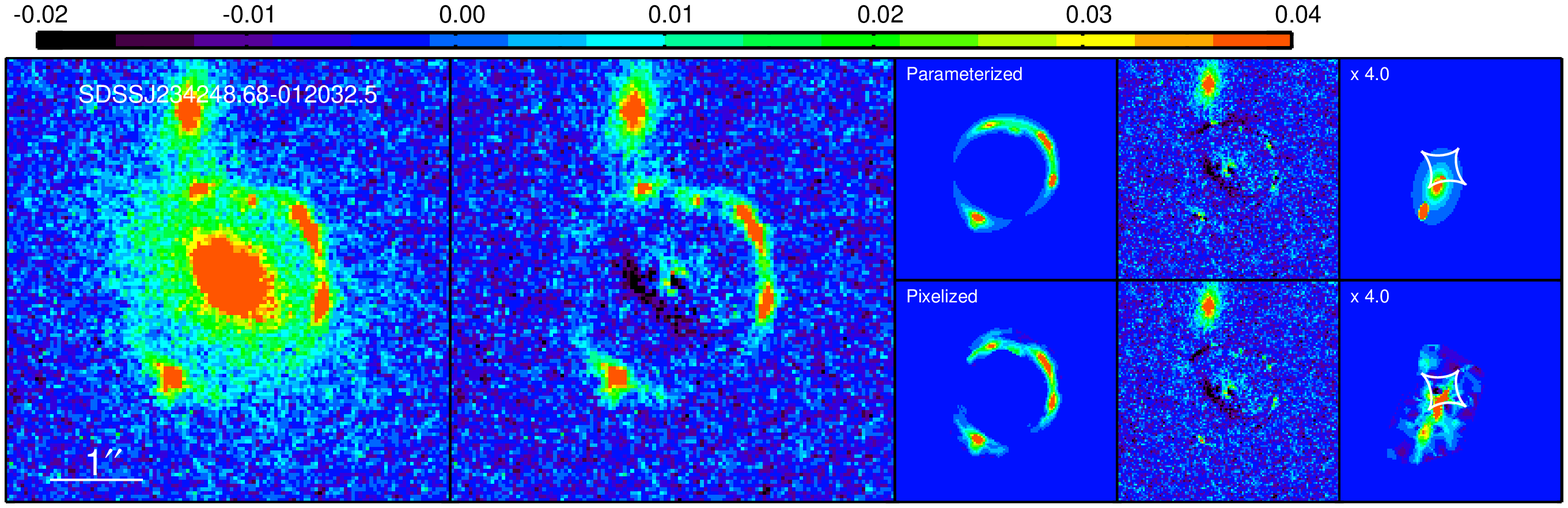}
\caption{{\label{fig:models_5}} \textit{Continued}}
\end{figure*}

\begin{figure*} 
\includegraphics[width=0.9\textwidth]{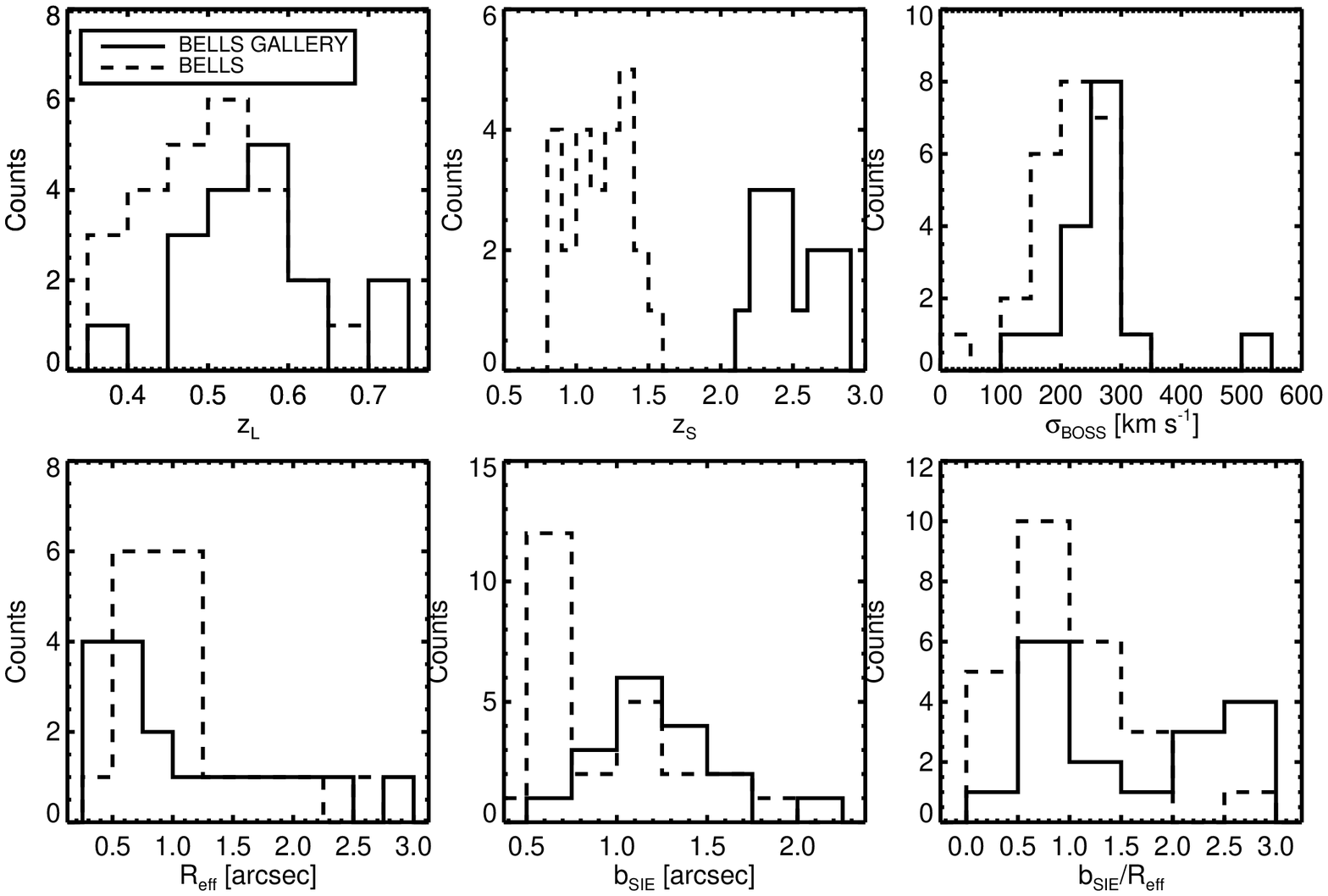}
\caption{\label{fig:bells_vs_gallery}
Distributions of the lens redshift, source redshift, stellar velocity dispersion, effective radius, Einstein radius, and the ratio of Einstein radius to effective radius for the BELLS (dashed histograms) and the BELLS GALLERY (solid histograms) lens samples.}
\end{figure*}

Dust extinction, especially differential extinction, in the lens galaxy could 
affect the lens modeling result. In addition, imperfect foreground-light 
subtraction may also introduce an equivalent effect. Therefore, 
we perform a simple test on the effects of differential extinction on the inferred 
SIE parameters using mock lenses with a double-image configuration. 
More specifically, we generate a double-image lens system with one image very 
close to the lens galaxy (analogous to J085621.59$+$201040.5). Then we perform the 
same lens modeling on this mock lens with the close-in image gradually dimmed, 
and compare the recovered parameter values to the input values. We find that for 
this mock lens, the Einstein radius is still well recovered with 
$\sim 7\%$ accuracy for a differential extinction of up to 1 mag. 
The SIE mass center shifts toward the close-in image in order to 
match its effectively lower magnification. The shift can reach as large as 
0\farcs05$-$ 0\farcs2 for a differential extinction of 0.5 mag. 

Although we cannot determine the dust (differential) extinction levels for 
the \Survey{} lens systems given the single-band \textsl{HST} imaging observations, 
a study by \citet{Eliasdottir06} shows that the mean differential extinction 
for their full lensing galaxy sample is only $0.33 \pm 0.03$ mag, a level that 
will not affect the fitting parameters significantly according to our test. 
In addition, our lenses are all massive ETGs, so we expect the amount of 
differential extinction to be relatively small 
\citep[e.g.,][]{Falco99, Kochanek99, Eliasdottir06, Vegetti12, More16}.

Table~\ref{tb:tb_source} provides the best-fit parameters for each individual 
source component for the 17 \Survey{} grade-A lenses. 
The quoted uncertainty in the effective radius is the statistical 
uncertainty, and the uncertainty in the apparent source magnitude is 
a combination in quadrature of the statistical uncertainty and a 0.1 mag 
systematic uncertainty due to foreground subtraction 
as suggested in \citet{Marshall07} and \citet{Shu16a}. 
Note that some source components are not well constrained with large uncertainties. 
An MCMC approach is demanded for a better source structure recovery.
Most of the systems require multiple source components with typical 
projected separations of 500 pc-2 kpc. The effective radii of the source 
components are as small as 0\farcs0037, or equivalently 31 pc. 
The best-fit magnification-corrected, rest-frame, far-UV, apparent AB magnitudes of 
the source components range from 29.6 to 24.2. 
Such resolution and sensitivity cannot be 
achieved without the aid of the average $\sim 13 \times$ lensing magnification. 
As a comparison, direct observations of unlensed LAEs at similar 
redshifts can only reach a UV magnitude limit of $~\sim 25-27$ 
\citep[e.g.,][]{Shapley03, Bond12, Kobayashi16}. 

Figure~\ref{fig:models_1} displays the results for both the parameterized 
and pixelized source models for the 17 grade-A lenses. For most of the 
lenses, smooth models obtained with the nonlinear fitting approach are able to 
explain the observational data down to the noise level, and the reduced $\chi^2$ 
values are close to unity (Table~\ref{tb:tb2}). 
There are bright residuals with significances greater than 10$\sigma$ 
in some lenses (e.g. J020121.39$+$322829.6, 
J075523.52$+$344539.5, and J091859.21$+$510452.5). These could be a sign for 
the presence of dark substructures, or caused by other artifacts such as 
an inappropriate PSF or foreground subtraction. 
A detailed analysis considering beyond the smooth lens model for these lens systems 
and the other \Survey{} grade-A lenses is deferred to a future paper.
The parameterized and pixelized source models agree reasonably well. 
The regularization level for the pixelized source model is chosen such that 
the resulting $\chi^2$ value is comparable to that of the parameterized 
source model. By design, the pixelized source model is optimized to capture 
the brightest pixels. Artificial, fragmentary structures are seen toward the 
lower surface brightness edges. 

\section{Discussion}

The lens galaxies in the \Survey{} sample are, by selection, similar in many ways to 
those in the BELLS sample. 
Figure~\ref{fig:bells_vs_gallery} compares some of the properties of the two samples. 
In particular, the lens redshift distributions are similar, and the typical sizes 
of the lens galaxies in the two samples 
as characterized by the median effective radii are $0\farcs98$ (\Survey{}) 
and $1\farcs00$ (BELLS), respectively. On the other hand, 
the \Survey{} lenses are relatively more massive than BELLS lenses 
with an average stellar velocity dispersion of 272 km s$^{-1}$ as compared to 
208 km s$^{-1}$. 
Combined with the higher source redshifts of the \Survey{} sample, 
the Einstein radii of the \Survey{} lenses are generally larger. 
The median Einstein radius of the 17 \Survey{} grade-A lenses is $1\farcs22$, 
while the median Einstein radius of the 25 BELLS grade-A lenses is $0\farcs75$ 
\citep{Brownstein12}. 
Consequently, the ratio of Einstein radius to effective radius for the 
\Survey{} grade-A lenses has a median of $1.37$, and $41\%$ (7/17) of 
the ratios are larger than two. 
These numbers are $0.80$ and $4\%$ (1/25) for the BELLS grade-A lenses.
As strong lensing provides an accurate estimate of the total mass 
within the Einstein radius, a joint analysis of the BELLS and \Survey{} lenses 
will strongly constrain the radius evolution of the mass profile in 
massive ETGs \citep[e.g.,][]{Rusin03, Oguri14}. 

The scientific motivations of the \Survey{} survey are to search for 
dark substructures within galaxies and extend the mass detection threshold 
using the intrinsically compact, high-redshift LAEs. 
As suggested by Figure~\ref{fig:models_1}, smooth (SIE plus external shear) 
lens models only accounting for the luminous components can already explain 
the observational data {of most of the lens systems} well. 
However, this only indicates the absence of substructures near the LOS 
to the lensed images with masses high enough to generate 
visually observable signals. In fact, previous dark substructure 
detections based on the lensed image surface brightness data are mostly confirmed 
statistically by comparing Bayesian evidences 
\citep[e.g.,][]{Vegetti10, Vegetti12, Fadely12, Hezaveh16}. 
Thus, the smooth lens models presented in this paper are only 
an important first step in searching for dark substructures in the \Survey{} lens 
sample. Further investigations under a Bayesian framework are deferred to future 
papers. 

Besides the lens galaxies, the lensed sources --- high-redshift LAEs --- 
are of great 
interest. LAEs are believed to be young, low-mass, and highly star-forming 
galaxies. They compose an important piece in our understanding of galaxy evolution 
scenarios 
\citep[e.g.,][]{Venemans05, Bond12, Ciardullo12, Ao15, Finkelstein15, 
Hathi16, Song16, Zheng16}. 
They can also be used as a probe of high-redshift 
circumgalactic and interstellar media \citep[e.g.,][]{Miralda-Escude98, Malhotra04, 
Zheng10, Zheng11b, Zheng14} and as tracers of high-redshift large-scale structures 
\citep[e.g.,][]{Hill08, Tamura09, Zheng11a}. 
In particular, the \textsl{HST} F606W filter covers the rest-frame 
far ultraviolet (UV) emission from the background LAEs. Previous studies of unlensed 
LAEs over a wide redshift range ($0.03 < z_{\rm LAE} < 7.0$) have shown that 
their far UV morphologies are sometimes ($\sim 20\%-50\%$) clumpy and/or irregular 
as a result of in situ or merger-triggered star formation 
\citep[e.g.,][]{Bond09, Bond12, Jiang13, Hayes14, Guaita15, Kobayashi16}.
However, as we have discussed in \citet{BELLSIII}, direct observations of LAEs 
are limited to the most luminous examples and 
any explorations of their ``fine'' structures 
on sub-kiloparsec scales are challenging and sometimes impossible. 
As shown in Table~\ref{tb:tb_source} and Figure~\ref{fig:models_1}, 
by combining the superb angular resolution of \textsl{HST} with 
the average $\sim 13 \times$ magnification of 
these lenses, we are able to resolve individual, faint star-forming 
knots in LAEs at redshifts from two to three smaller than $\sim100$ pc. 
A detailed study of the properties of the lensed LAEs and comparisons with 
their unlensed counterparts are deferred to a forthcoming paper. 
Furthermore, infrared and sub-millimeter spectroscopic 
follow-up observations of the highly magnified \Survey{} lens sample targeted at 
the rest-frame atomic and molecular lines in the LAEs will permit exploration of 
the interstellar medium (ISM) and circumgalactic medium (CGM) 
of these high-redshift galaxies. 

\section{Conclusions}

We present \textsl{HST} WFC3 F606W-band imaging observations of 21 
galaxy-Ly$\alpha$ strong gravitational lens candidates. 
The sample, known as the \Survey{} sample, was 
spectroscopically selected from almost 1.5 million galaxy spectra in 
the final data release of the BOSS survey in SDSS-III. 
The foreground-lens galaxies are at a typical redshift of $0.55$ and the 
Ly$\alpha$ emission comes from redshifts from two to three. 

The \textsl{HST} data are fully reduced and analyzed by the custom-built tools 
{\tt ACSPROC} and {\tt lfit\_gui}, respectively. 
After modeling the systems with smooth lens models consisting of SIE mass 
distributions for the luminous components and an external shear, 
the main findings are as follows:
\begin{enumerate}
	\item Seventeen systems are confirmed to be grade-A lenses including 8 with 
			extended arcs, 3 with quadruple images, and 6 with double images. 
			Three systems are singly imaged 
			non-lenses, while the remaining system is a temporary ``maybe'' with 
			complex structures that are hard to interpret based solely on 
			single-band data. Considering the typical $50\%$ success rate 
			in the previous surveys with similar selection techniques, 
			another $\sim 70$ galaxy-LAE strong lenses are expected among 
			the remaining 166 lens candidates in the parent sample.
	\item We demonstrate that different foreground-light subtraction schemes can 
			lead to different model parameter estimations particularly for 
			two image systems where the fractional differences in the inferred 
			Einstein radii are $2\%-3\%$. This result highlights the need of 
			performing foreground-light subtraction jointly with the lens modeling.
	\item Because of the much higher source redshifts and more massive lens 
			galaxies, the Einstein radii of the \Survey{} lenses are generally 
			larger than those of the BELLS lenses, while the lens galaxy sizes 
			are comparable. As a result, the combination of the \Survey{} and 
			BELLS lenses can constrain any radius evolution of the mass profile 
			in massive ETGs. 
	\item The smooth lens models seem to be adequate for explaining the observed 
			imaging data for most of the 17 \Survey{} grade-A lenses 
			as shown in Figure~\ref{fig:models_1}. Although this suggests the 
			absence of massive dark substructures near the lensing features, 
			a thorough exploration of substructures in a statistical manner 
			is required.
	\item The average lensing magnifications of the background LAEs are found to be 
			$4-26$. The LAEs are thus resolved into individual star-forming 
			knots of a wide range of properties. They have characteristic sizes from 
			less than 100 pc to several kiloparsecs, rest-frame, far-UV, apparent AB 
			magnitudes from 29.6 to 24.2, and typical projected separations of 
			500 pc to 2 kpc. Further follow-up spectroscopic observations 
			will reveal the ISM and CGM properties of these high-redshift galaxies. 
\end{enumerate}

\acknowledgments

We thank the anonymous referee for helpful comments. This work has been partially supported by the Strategic Priority Research Program ``The Emergence of Cosmological Structures'' of the Chinese Academy of Sciences Grant No. XDB09000000 and by the National Natural Science Foundation of China (NSFC) under grant numbers 11333003, 11390372 (YS and SM), and 11603032 (YS). C.S.K. is partially supported by NSF grant AST-1515876. The work of M.O. was supported in part by World Premier International Research Center Initiative (WPI Initiative), MEXT, Japan, and JSPS KAKENHI Grant Number 26800093 and 15H05892. Z.Z. is partially supported by NASA grant NNX14AC89G and NSF grant AST-1208891. B.M. acknowledges support from NSF-1313302. 

Support for program \# 14189 was provided by NASA through a grant from the Space Telescope Science Institute, which is operated by the Association of Universities for Research in Astronomy, Inc., under NASA contract NAS 5-26555. 

Funding for SDSS-III was provided by the Alfred P. Sloan Foundation, the Participating Institutions, the National Science Foundation, and the U.S. Department of Energy Office of Science. The SDSS-III website is http://www.sdss3.org/.

SDSS-III was managed by the Astrophysical Research Consortium for the Participating Institutions of the SDSS-III Collaboration including the University of Arizona, the Brazilian Participation Group, Brookhaven National Laboratory, Carnegie Mellon University, University of Florida, the French Participation Group, the German Participation Group, Harvard University, the Instituto de Astrofisica de Canarias, the Michigan State/Notre Dame/JINA Participation Group, Johns Hopkins University, Lawrence Berkeley National Laboratory, Max Planck Institute for Astrophysics, Max Planck Institute for Extraterrestrial Physics, New Mexico State University, New York University, Ohio State University, Pennsylvania State University, University of Portsmouth, Princeton University, the Spanish Participation Group, University of Tokyo, University of Utah, Vanderbilt University, University of Virginia, University of Washington, and Yale University.


\end{document}